\documentclass[prx,twocolumn,amsmath,amssymb,aps,reprint,nofootinbib]{revtex4-2}

\usepackage{siunitx}
\usepackage{caption}
\usepackage{subcaption}
\usepackage{graphicx}
\usepackage{dcolumn}
\usepackage{bm}
\usepackage{braket}
\usepackage{float}
\usepackage{multirow}
\usepackage{booktabs}
\usepackage[table,xcdraw]{xcolor}
\usepackage{adjustbox}
\usepackage{hyperref}
\usepackage[capitalise]{cleveref}

\begin{document}

\newcommand{\code}[3]{[\![#1,#2,#3]\!]}

\preprint{APS/123-QED}
\title{Individually Addressed Logical Pauli Measurements \\ by QLDPC Surgery on a Trapped Ion Quantum Computer}

\author{Surabhi Luthra$^\ddagger$}
\email{zcaplut@ucl.ac.uk}
\affiliation{Department of Physics \& Astronomy, University College London, London WC1E 6BT, United Kingdom}

\author{Alex Owens$^\ddagger$}
\email{Alexander.Owens@STFC.ac.uk}
\affiliation{National Quantum Computing Centre, Harwell Campus, Didcot, United Kingdom}

\date{\today}

\def\thefootnote{$\ddagger$}\footnotetext{These authors contributed equally to this work}\def\thefootnote{\arabic{footnote}}

\begin{abstract}
Quantum low-density parity check (QLDPC) codes may lead to order-of-magnitude reductions in the qubit overhead required for fault-tolerant quantum computation. Achieving this improved overhead requires logical operations that selectively address individual logical qubits encoded within multi-qubit code blocks. Recently proposed architectures achieve this through QLDPC surgery, in which non-destructive Pauli measurements are performed using ancillary ``gadget" systems transiently entangled with the code block, generalising lattice surgery. Here we implement QLDPC surgery on a [[10,2,3]] generalised bicycle code using Quantinuum's H2-2 trapped ion processor. A single-round surgery measurement using seven gadget qubits measures the $\bar{X}_1$ operator with a 0.72 [0.32, 1.41]\% (95\% credible interval) error rate and agrees with subsequent transversal measurement with 1.03 [0.60, 1.65]\% error. Decoding across three rounds of syndrome extraction suppresses the gadget measurement error, reducing the error rate to below 0.26\%. The syndrome after surgery confirms return to the code space and the unaddressed ``spectator'' qubit is preserved with error 0.21 [0.04, 0.67]\% during the three-round measurement. We further demonstrate composition of conjugate basis gadgets by sequentially measuring $\bar{X}_1$ then $\bar{Z}_2$, reusing the gadget qubits, and obtain a 1.12 [0.60, 1.94]\% error in the $\bar{Z}_2$ measurement. These results constitute a hardware validation of the generalised surgery primitive underlying Pauli-based computing for QLDPC architectures.
\end{abstract}

\maketitle

\section{\label{sec:intro}Introduction}
Fault-tolerant quantum computation based on the surface code requires hundreds to thousands of physical qubits per logical qubit, placing scientifically and commercially relevant problems beyond the reach of near-term hardware \cite{Fowler2012, Litinski2019, Gidney2025}. Certain quantum low-density parity check (QLDPC) codes reduce this overhead by encoding multiple qubits in a single code block while retaining bounded-weight parity checks \cite{Breuckmann2021, Bravyi2024} at the cost of requiring connectivity beyond nearest-neighbour. Recent architectural proposals built on such codes \cite{Webster2026Pinnacle, Yoder2025TourDeGross, Cain2026} project order-of-magnitude reductions in the physical qubit count required for applications such as factoring and quantum simulation. 

Realising these savings requires a method for selectively addressing logical measurements within a code block of more than one logical qubit. An approach extensively explored theoretically is QLDPC surgery (a generalisation of lattice surgery \cite{Horsman2012}). In QLDPC surgery the code block is dynamically merged with a gadget - an ancillary qubit system - such that the product of a set of low-weight check measurements on the gadget yields the eigenvalue of a chosen logical Pauli operator \cite{Cohen2022,Cross2025,Williamson2026,He2025Extractors}. Combined with a source of magic states, logical Pauli-product-measurement suffices for universal quantum computing in a model referred to as Pauli-based computing \cite{Litinski2019,Bravyi2016PBC}. A defining feature of the measurements required for Pauli-based computing is that they are non-destructive \cite{Bravyi2016PBC}, meaning that they preserve the projected eigenstate while keeping the system within the logical code space after measurement. In contrast, transversal measurement of a logical qubit is inherently destructive as it collapses the encoded state and projects the system out of the code space. 

While this work focusses on surgery for general QLDPC codes, we note a complementary approach for fault-tolerant logic based on designing high-rate codes with rich automorphism groups. Examples of this are the subsystem hypergraph product simplex QLDPC codes \cite{Malcolm2026shyps} or concatenated symplectic double codes \cite{Berthusen2026simplelogical}, which exploit reconfigurable or high non-local connectivity and fold-transversal gates to realise logical operations, with the potential to avoid the use of ancilla systems for Clifford operation altogether. The fold-transversal approach is made universal through the injection of magic states via chain-map CNOT gates \cite{benhemou2026automatedlogicalcliffordgadgets}, or indeed surgery, with the chain map CNOT approach recently demonstrated at beyond break-even error \cite{berthusen2026experimentalvalidationcompactfaulttolerant}. The connectivity required for surgery is defined by a fixed Tanner graph which may be known in advance, permitting physical implementations with fixed couplers and limited long-range connectivity \cite{Yoder2025TourDeGross,He2025Extractors}. Compared to hardware with reconfigurable couplings (e.g. atomic qubits through transport or optical switches), fixed connectivities incur a greater physical qubit overhead for surgery as dynamic connectivity allows reuse of the same qubits to define new gadgets where fixed couplers require a universal gadget system implemented at the device design stage.  Generalised surgery provides addressable, non-destructive logical Pauli measurement for general QLDPC codes with low ancilla overhead \cite{Webster2025Gadgets}, and architectures have been proposed combining elements of both approaches \cite{Webster2026Pinnacle}. 

Research into generalised surgery continues to develop rapidly, for example tailoring techniques to specific QLDPC codes \cite{Blue2026,Zheng2026,Bhardwaj2026, Gu2026}. Recent experiments have reported the first demonstrations of QLDPC memory with beyond nearest-neighbour connectivity in trapped ions \cite{Hong2024,Tham2026}, superconducting \cite{Wang2026} and neutral atom platforms \cite{Atom2026}. Surgery demonstrations have so far been limited to lattice surgery where code blocks each encode a single logical qubit \cite{Erhard2021lattice_entangle,Besedin2026lattice_SC,wang2026superconductingsurfacecodeprocessorlatticesurgery}.
To the best of our knowledge, measuring individual logical qubits in a multi-qubit code block using QLDPC surgery has not been realised in hardware. In this work, we demonstrate individually addressed logical Pauli measurement via QLDPC surgery on a \code{10}{2}{3} generalised bicycle code, using the Quantinuum H2-2 trapped ion processor \cite{QuantinuumH2}. We focus on the explicit low-overhead gadget construction presented in Ref. \cite{Webster2025Gadgets} which is applicable to arbitrary QLDPC codes and forms the basis of the processing blocks in the Pinnacle architecture \cite{Webster2026Pinnacle}. 

We first establish that a single round of gadget measurement correctly extracts the eigenvalue of the targeted logical operator across a set of logical input states (Section \ref{sec:gadget-operation}) and that this value agrees with subsequent transversal measurement. We then demonstrate repetition of the gadget measurement across $R=3$ rounds of syndrome extraction suppresses measurement error, showing the mechanism by which the protocol is made fault-tolerant (Section \ref{sec:ft}). We verify the gadget measurement preserves the code state by analysis of the effect of the gadget on the syndrome and we demonstrate the preservation of the spectator logical qubit through transversal measurement. Finally we sequentially measure $\bar{X}_1$ then $\bar{Z}_2$ by surgery using the same gadget qubits.

\section{Code and Gadget Construction\label{sec:code-construction}}

\subsection{The \code{10}{2}{3} Generalised Bicycle Code}

Generalised bicycle (GB) codes are a family of Calderbank–Shor–Steane (CSS) QLDPC codes parametrised by a lift parameter $l$ and a pair of binary polynomials $a(x), b(x) \in \mathbb{F}_2[x]/(x^l-1)$ \cite{Lin2025,Wang2022,Kovalev2013}. These polynomials define circulant parity checks from which the $X$- and $Z$-type stabilisers are constructed.

\begin{figure}[h]
\centering\includegraphics[width=\linewidth]{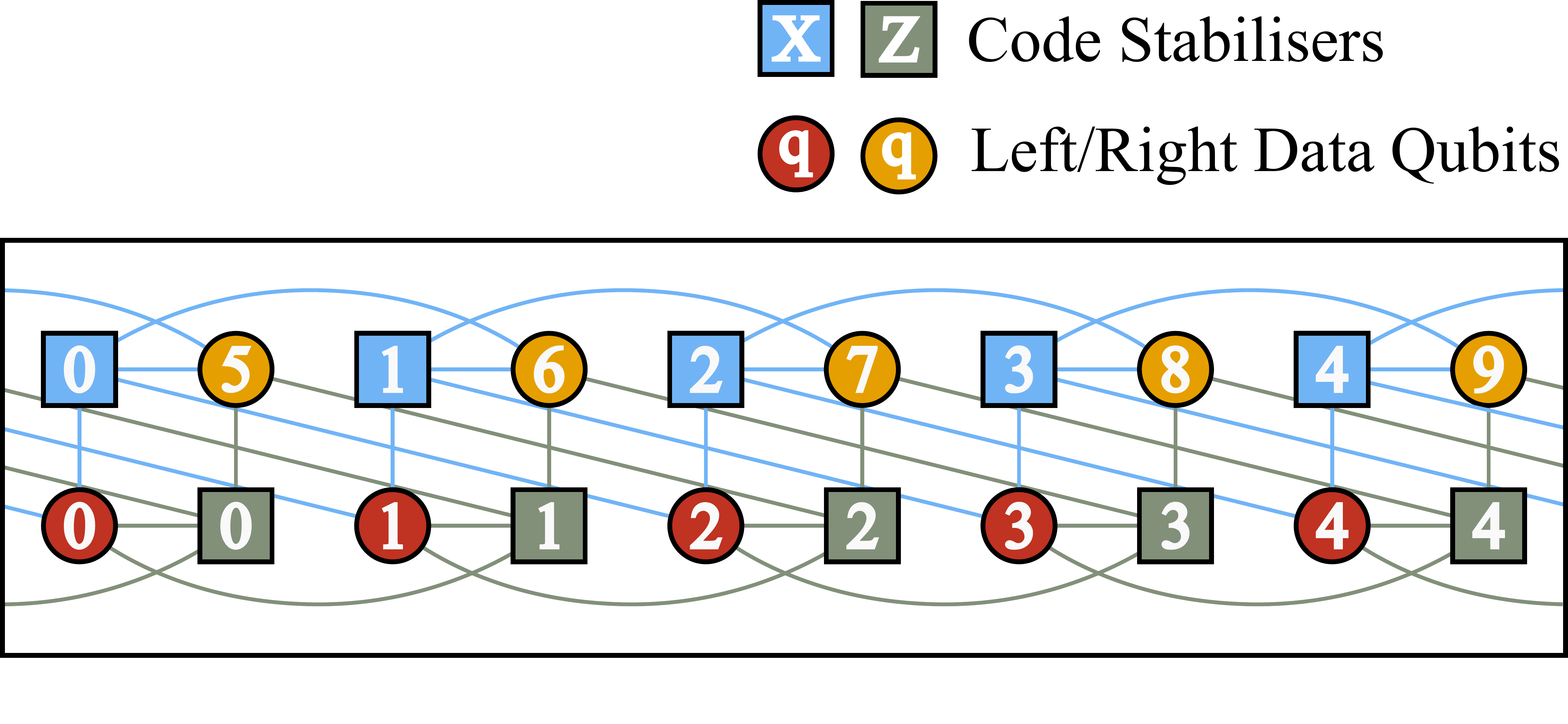}
    \caption{The \code{10}{2}{3} generalised bicycle code construction with $l=5$, $a(x)=1+x^2$, and $b(x)=1+x$. For readability we omit all but the numeric label from the nodes, encoding $X$- and $Z$-type checks and left- and right-block data qubits in the colour and shape of the nodes as indicated in the legend. \label{fig:code}}
    \end{figure}

In this work, we consider the GB code specified by \begin{align} l=5,\qquad a(x)=1+x^2,\qquad b(x)=1+x. \end{align} The resulting \code{10}{2}{3} code encodes two logical qubits into ten physical data qubits arranged into left and right blocks of five qubits each, as shown in Fig.~\ref{fig:code}. The code has distance $3$, and all stabiliser generators have weight $4$, reflecting the two non-zero coefficients in each polynomial.

The logical operators of this code admit minimum weight representations: 
\begin{align}
    \bar{X}_1 &= X_{q_2} X_{q_3} X_{q_7}, \label{eqn:X1 rep} \\
    \bar{X}_2 &= X_{q_1} X_{q_7} X_{q_9}, \label{eqn:X2 rep} \\
    \bar{Z}_1 &= Z_{q_2} Z_{q_4} Z_{q_5}, \label{eqn:Z1 rep} \\
    \bar{Z}_2 &= Z_{q_2} Z_{q_6} Z_{q_7}, \label{eqn:Z2 rep}
\end{align}
where the data qubit indices $0-4$ correspond to the left block and $5-9$ correspond to the right block, as shown in \cref{fig:code}.

The \code{10}{2}{3} code can be equivalently constructed as a rotated toric code \cite{Arnault2025,rotated_toric}. However, for the purposes of this work, the \code{10}{2}{3} code serves as a minimal, non-trivial multi-qubit code block on which to demonstrate individually addressed non-destructive measurement by QLDPC surgery. The construction we demonstrate \cite{Webster2025Gadgets} is applicable to general QLDPC codes, including larger bicycle codes without a geometric description which we leave to future work. 

\subsection{Logical Measurement Gadgets\label{sec:gadget-construction}}

We construct a measurement gadget for a single logical Pauli operator following the method in Ref.  \cite{Webster2025Gadgets}. For an $X$-type logical operator $\mathbf{L}$ with support $\{q_i\}$, the gadget comprises three distinct sets of ancilla qubits: the gadget qubits (labelled as $\kappa$), the $X$-type gadget checks ($\chi$), and the $Z$-type gauge checks ($\gamma$). The configuration of the gadget is as follows:

\begin{itemize}
    \item \textbf{Gadget qubits ($\kappa$):} One gadget qubit $\kappa_j$ for each $Z$-type stabiliser $S_{Z,j}$ that overlaps with $\mathrm{supp}(\mathbf{L})$, with $\kappa_j$ subsequently appended to the support of $S_{Z,j}$.
    
    \item \textbf{$X$-type gadget checks ($\chi$):} One $X$-type gadget check $\chi_i$ for each data qubit $q_i \in \mathrm{supp}(\mathbf{L})$. This check is supported on $q_i$ and its associated gadget qubits, defined by the set $\{\kappa_j \mid q_i \in \mathrm{supp}(S_{Z,j})\}$.
    
    \item \textbf{$Z$-type gauge checks ($\gamma$):} A set of $|S_{\mathbf{L}}| - \mathrm{wt}(\mathbf{L}) + 1$ $Z$-type gauge-fixing checks $\gamma_m$ supported exclusively on the gadget qubits. The support of these checks must be chosen such that they form a basis for $\ker \left(H_{X,\mathrm{gadget}}\right)$.
\end{itemize}

The set of gadget qubits $\kappa$ act as a measurement interface for the code. The $X$-type gadget checks $\chi$  encode the logical observable  $\prod_i \chi_i = \mathbf{L}$  such that their parity gives the logical measurement outcome. The $Z$-type gauge-fixing checks $\gamma$ remove unwanted degrees of freedom introduced by the gadget qubits so any new operators are promoted to stabilisers of the merged code. 

\begin{figure}[h]
\centering
\includegraphics[width=0.9\linewidth]{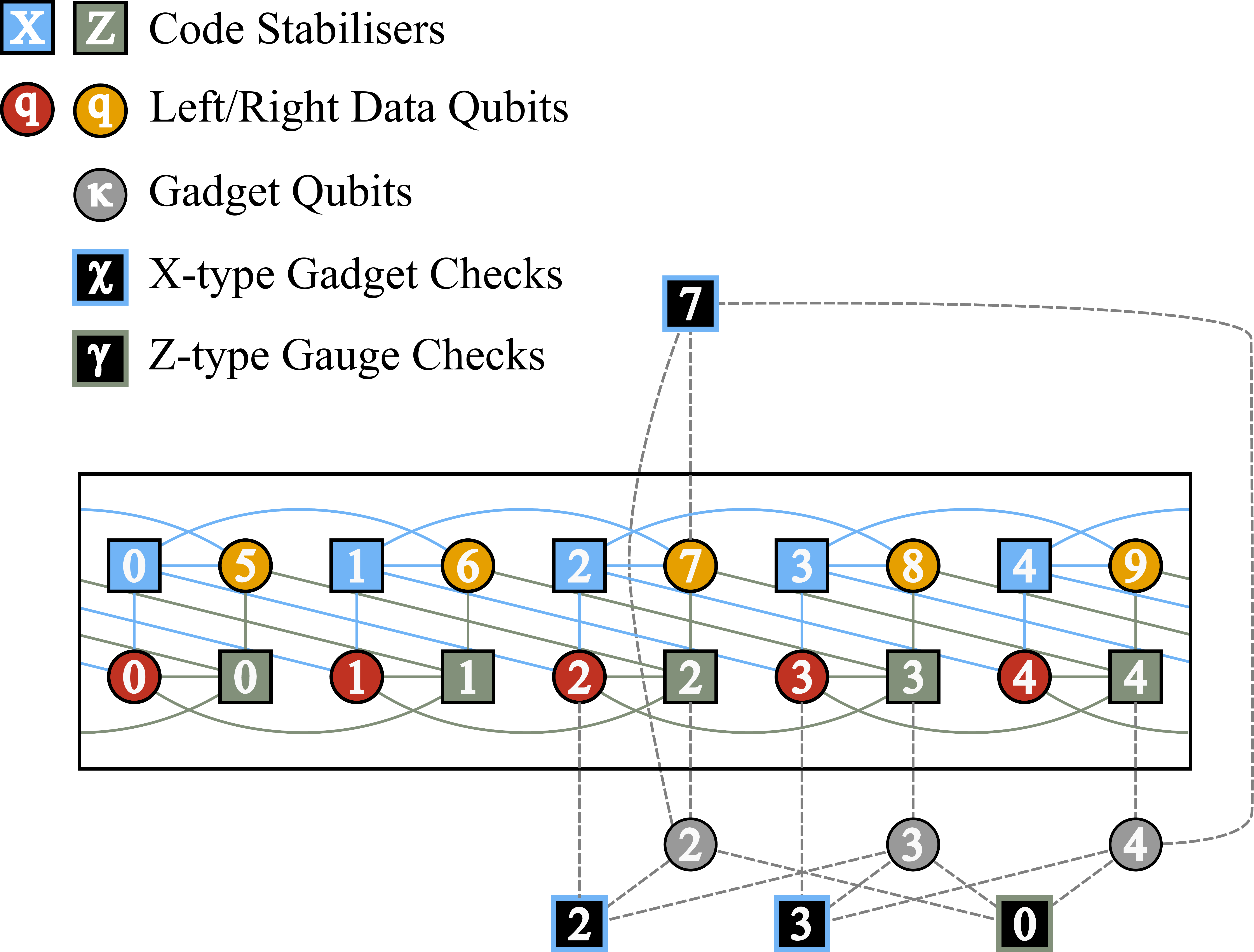}
\caption{\code{10}{2}{3} code merged with $\bar{X}_1$ measurement gadget. \label{fig:gadget}} 
\end{figure}

To demonstrate this explicitly for the \code{10}{2}{3} code, consider the measurement of the logical operator $\bar{X}_1$ with $\text{supp}(\bar{X_1}) = \{q_2, q_3, q_7\}$. Following the steps above, the measurement gadget for this operator consists of 7 ancillary qubits as outlined below. We illustrate the support of this gadget on the code in \cref{fig:gadget}.  

\begin{itemize}
    \item \textbf{Gadget qubits ($\kappa$):}  We first identify three $Z$-type stabilisers that overlap with the logical support of $\bar{X}_1$, namely $S_{Z,2}$, $S_{Z,3}$ and $S_{Z,4}$. Thus, we introduce three gadget qubits: $\kappa_2$, $\kappa_3$, and $\kappa_4$.
    \item \textbf{$X$-type gadget checks ($\chi$):} We require one $X$-type check for each of three data qubits in the logical support:
    \begin{itemize}
        \item $\chi_2$ is supported on data qubit $q_2$ and gadget qubits $\{\kappa_2, \kappa_3\}$ (since $q_2 \in \mathrm{supp}(S_{Z,2})$ and $\mathrm{supp}(S_{Z,3})$).
        \item $\chi_3$ is supported on data qubit $q_3$ and gadget qubits $\{\kappa_3, \kappa_4\}$ (since $q_3 \in \mathrm{supp}(S_{Z,3})$ and $\mathrm{supp}(S_{Z,4})$).
        \item $\chi_7$ is supported on data qubit $q_7$ and gadget qubits $\{\kappa_2, \kappa_4\}$ (since $q_7 \in \mathrm{supp}(S_{Z,2})$ and $\mathrm{supp}(S_{Z,4})$).
    \end{itemize}
        \item \textbf{$Z$-type gauge checks ($\gamma$):} We require $|S_{\bar{X}_1}| - \mathrm{wt}(\bar{X}_1) + 1 = 3 - 3 + 1 = 1$ gauge check. To find the basis of $\ker \left(H_{X,\mathrm{gadget}}\right)$, we observe that each gadget qubit appears in exactly two $X$-type checks. Therefore, a single $Z$-type check supported on all three gadget qubits, $\gamma_0 = Z(\{\kappa_2, \kappa_3, \kappa_4\})$, will have an even overlap with every $\chi$ check, satisfying the kernel condition.
\end{itemize}

We can see from this construction that 
\begin{align}
\prod_i \chi_i &= \chi_2 \chi_3 \chi_7 \\ &= (X_{q_2}X_{\kappa_2}X_{\kappa_3})(X_{q_3}X_{\kappa_3}X_{\kappa_4})(X_{q_7}X_{\kappa_2}X_{\kappa_4}) \\ &= X_{q_2}X_{q_3}X_{q_7} \\ &= \bar{X}_1.
\end{align}

While the $\gamma$ checks in the construction constrain the degrees of freedom associated with the cycles in the gadget, the distance of the merged code-gadget system can be reduced if the Tanner graph represented by $H_{X,\mathrm{gadget}}$ has insufficient expansion, quantified by the boundary Cheeger constant $h$ \cite{Cross2025}. Insufficient expansion can lead to low-weight representatives of the logical operators and a drop in distance while the merged code retains the distance of the bare code for $h\geq1$. The gadgets constructed in this work have a boundary Cheeger constant of $h=2$, retaining distance 3. 

\begin{figure}[h]
\centering
\includegraphics[width=0.9\linewidth]{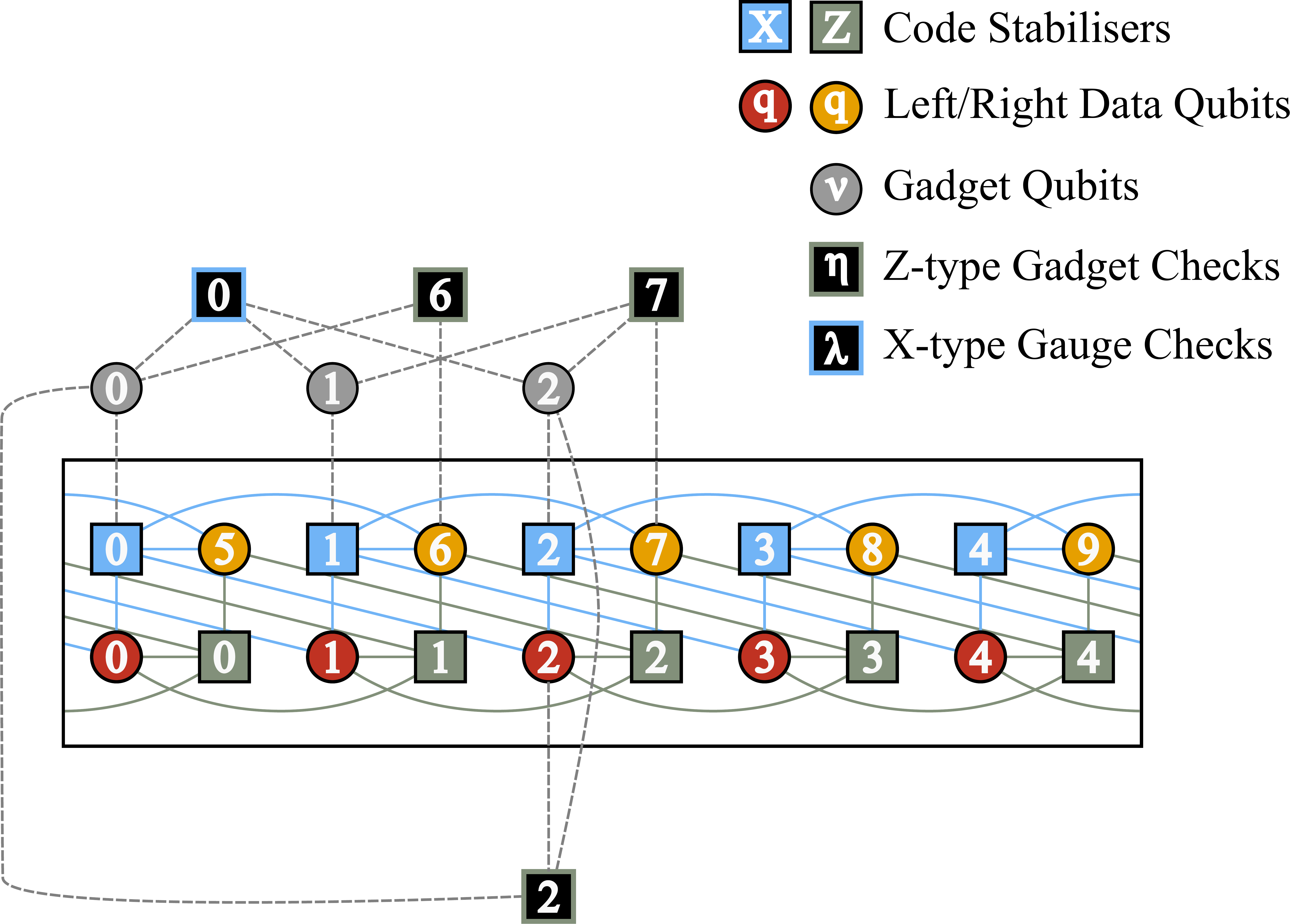}
\caption{\code{10}{2}{3} code merged with $\bar{Z}_2$ measurement gadget. Here we label the gadget qubits as $\nu$, $Z$-type gadget checks as $\eta$ and $X$-type gauge checks as $\lambda$ for distinction with the $X$-type gadget in \cref{fig:gadget}. \label{fig:gadget-Z2}} 
\end{figure}

We can construct measurement gadgets of each of the four logical operators $\{ \bar{X}_1, \bar{X}_2, \bar{Z}_1, \bar{Z}_2\}$ of this code in a similar manner, where we swap the roles of $X$ and $Z$ basis in the construction for the $Z$-type logical operators as shown in \cref{fig:gadget-Z2}.

\section{Experimental Methods\label{sec:methods}}
\begin{figure*}[t]
\centering
\begin{subfigure}{\textwidth}
\centering
\includegraphics[width=\textwidth]{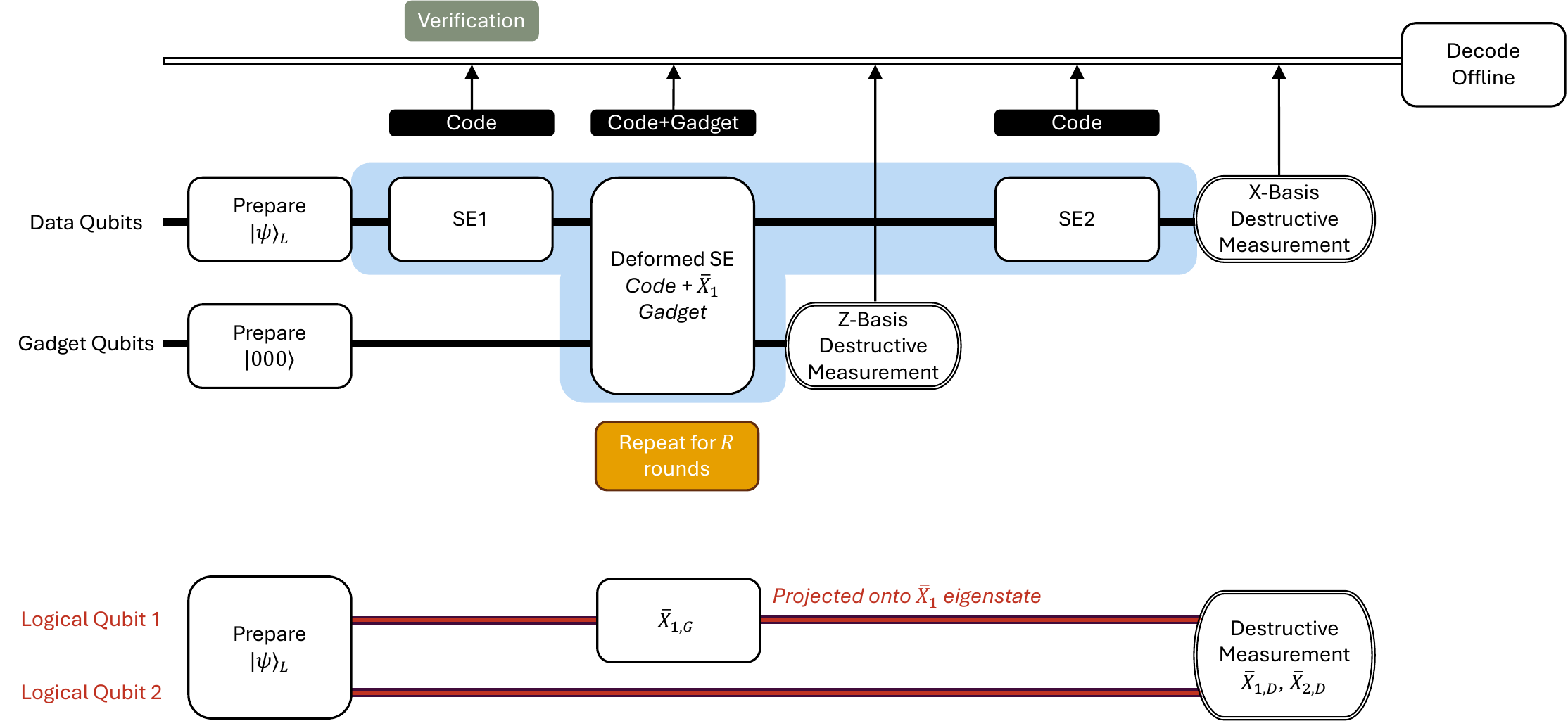}
\caption{\label{fig:X1-circuit}} 
\end{subfigure}

\begin{subfigure}{\textwidth}
\centering
\includegraphics[width=\textwidth]{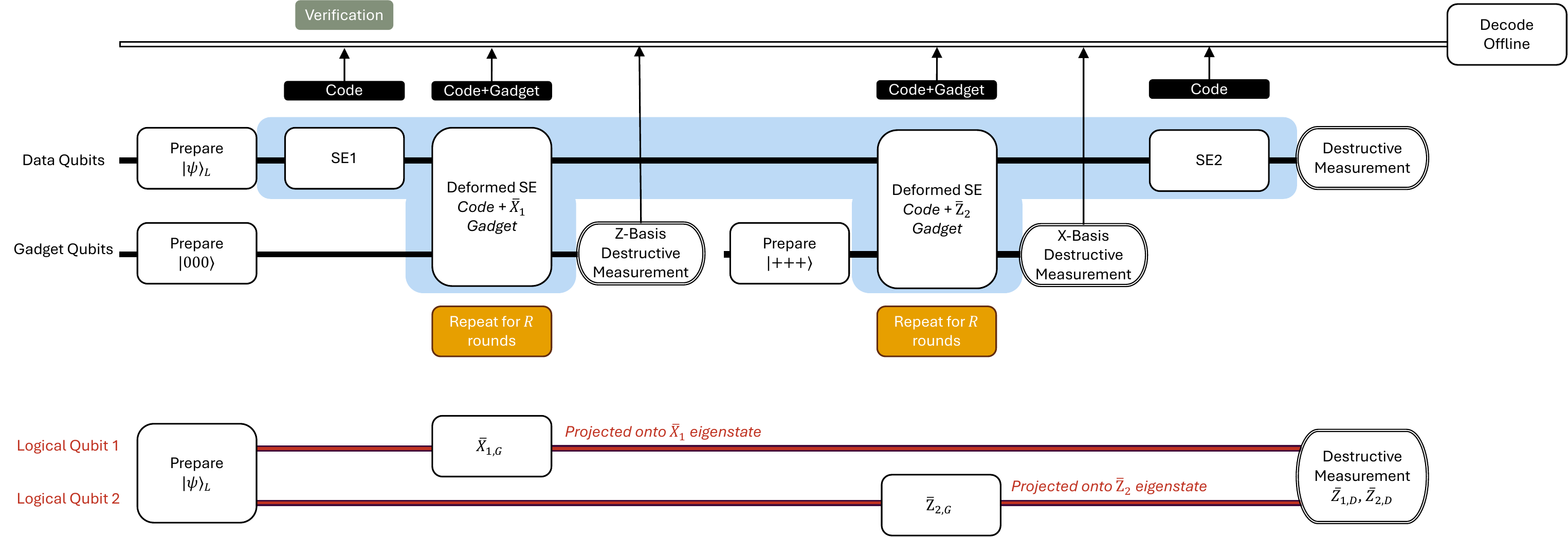}
\caption{\label{fig:X1-Z2-circuit}} 
\end{subfigure}
\caption{Experiment flow at the physical (upper) and logical (lower) level. Ancilla not pictured. Blue shading represents which qubits comprise the code block at each time-step, representing the code deformation. (a) Single-gadget circuit. The logical state $\ket{\psi}_L$ and is prepared on the data qubits and the gadget qubits $\kappa$ are initialised in $\ket{0}$. The preparation of $\ket{\psi}_L$ is verified by trivial syndrome in the first bare code syndrome extraction SE1. The gadget and code are merged and the deformed codes syndrome extraction is repeated $R$ times. The gadget is detached from the code by measuring the $\kappa$ qubits in the $Z$ basis. After the system is returned to the \code{10}{2}{3} code there is one final bare code syndrome extraction SE2, followed by terminal transversal measurement of the data qubits in $X$ ($\bar{X}_{1,D}$,$\bar{X}_{2,D}$). (b) Sequential two-gadget circuit. As in a) until the $\bar{X}_{1}$ gadget is measured out, then the same gadget qubits are re-used prepared in $\ket{+}$ as the $\nu$ qubits of a $Z$-type gadget. The $Z$ gadget is merged with the code, with the deformed code syndrome extraction repeated for $R$ rounds (only $R=1$ reported in this work), before the $\nu$ qubits are measured in $X$, returning the system to the \code{10}{2}{3} code. A final syndrome extraction on the bare code (SE2) is followed by terminal transversal measurement in $Z$ ($\bar{Z}_{1,D}$,$\bar{Z}_{2,D}$). }
\end{figure*}

In this section, we outline the experiments that we performed to demonstrate various properties of the measurement gadget on hardware. We create three distinct gadget circuits: An $\bar{X}_1$ gadget with a single round of merged code-gadget syndrome extraction which we label XR1 (Fig. \ref{fig:X1-circuit}), an $\bar{X}_1$ gadget circuit with the merged code-gadget syndrome extraction repeated for 3 rounds labelled XR3, and a circuit in which we measure $\bar{X}_1$ then $\bar{Z}_2$ (both with a single round gadget) sequentially using the same gadget qubits, which we label SEQ (Fig. \ref{fig:X1-Z2-circuit}). In addition, we also conduct a memory experiment on the code without a gadget, with 5 repeated syndrome extraction cycles which we label MEM. \cref{tab:experiments} summarises the experiments conducted in this work.

\begin{table}[h]
    \centering
    \renewcommand{\arraystretch}{1.5} 
    \begin{tabular}{|c|c|c|c|c|c|}
        \hline
        \textbf{Circuit} & \textbf{Data} & \textbf{Gadget} & \textbf{Gadget} & \textbf{Meas.} & \textbf{Shots} \\
        & \textbf{Prep.} & & \textbf{Rounds ($R$)} & \textbf{Basis} & \\
         \hline
         XR1 & $\lvert++\rangle_L$ & $\bar{X}_1$ & 1 & $X$ & 500 \\
         XR1 & $\lvert--\rangle_L$ & $\bar{X}_1$ & 1 &  $X$ & 500 \\
         XR1 & $\lvert00\rangle_L$ & $\bar{X}_1$ & 1 & $X$ & 250 \\
         XR1 & $\lvert11\rangle_L$ & $\bar{X}_1$ & 1 & $X$ & 250 \\
         XR3 & $\lvert++\rangle_L$ & $\bar{X}_1$ & 3 & $X$ & 1000 \\
         SEQ & $\lvert00\rangle_L$ & $\bar{X}_1$, $\bar{Z}_2$ & 1 (each gadget) & $Z$ & 500 \\
         SEQ & $\lvert11\rangle_L$ & $\bar{X}_1$, $\bar{Z}_2$ & 1 (each gadget) & $Z$ & 500 \\
         MEM & $\lvert++\rangle_L$ & None & 0 & $X$ & 500 \\
         \hline
    \end{tabular}
    \caption{Summary of experiments. Each experiment consists of data qubits prepared in a logical state and measured in the appropriate basis. Each gadget circuit (XR1, XR3, SEQ) also contains a bare code syndrome extraction cycle before and after the gadget rounds labelled SE1 and SE2 respectively. The memory experiment (MEM) consists of 5 bare syndrome extraction cycles and no gadget. \label{tab:experiments}}
\end{table}

\subsection{Gadget Circuit Structure\label{sec:circuit}}

Each experimental shot proceeds as follows. A logical state of the two encoded qubits is prepared then verified on trivial outcome of the initial bare code syndrome extraction round (SE1). The gadget qubits are initialised in $\ket{0}$ at the start of the experiment. Following SE1, the combined gadget and code system syndrome extraction is repeated $R$ times. This extends the $Z$-type stabilisers $$S_{Z,j} \to S_{Z,j} \otimes Z{\kappa_j}.$$
The $\kappa_j$ qubits are then measured in the $Z$ basis, returning the code block to the bare code, up to a Pauli frame correction determined by the $\kappa$ outcomes. 

We follow the procedure of Ref. \cite{Cross2025} to perform this \textit{fusion correction}. Specifically, the set of stabilisers that interact with the measurement gadget $\{S_{Z, 2}, S_{Z, 3}, S_{Z, 4}\}$ are modified when the code is merged due to additional support on the $\kappa$  qubits. When $\kappa$ qubits are subsequently measured to disentangle the gadget, their measurement outcomes determine the signs of the corresponding code stabilisers. For example, the stabiliser $S_{Z, 2}$ undergoes the following transformation during this process: $$S_{Z, 2} \to S_{Z, 2} \otimes Z_{\kappa_2} \to m(\kappa_2) \cdot S_{Z,2}.$$ Similarly, the gauge check $\gamma_0$ is also transformed to $$m(\kappa_2) m(\kappa_3) m(\kappa_4) \cdot \gamma_0.$$ Finally, we perform another syndrome extraction round (SE2) on the code before destructively measuring the data qubits. In Appendix~\ref{app:a}, we present the detector matching graph for the XR3 experiment, inspired by Ref. \cite{Cross2025}. 

For the two-gadget circuit, we repeat the procedure for the single-gadget circuit, except we now introduce another round of combined code and gadget syndrome extraction for the second $\bar{Z}_2$ gadget. This requires measuring the $\kappa_j$ qubits to separate the gadget system and code with appropriate fusion corrections as above, then reinitialising the gadget to perform the second surgery. After the second gadget is performed and measured, the $X$-type stabilisers coupled with the $\bar{Z}_2$ gadget qubits require fusions corrections analogous to those for the $\bar{X}_1$ gadget.  

We schedule syndrome extraction for the code stabilisers using the procedure of Ref. \cite{Lin2025}. For the \code{10}{2}{3} code, this simplifies to the standard N--Z scheduling across four time steps. The same scheduling construction applies more generally to larger GB codes and therefore provides a route to scaling the present circuits in future work. The gates required to measure the additional gadget checks $\chi$ and $\gamma$ are scheduled over three further time steps. 

The merged codes associated with both gadgets in \cref{fig:gadget,fig:gadget-Z2} have code distance $d=3$ by construction (Sec. \ref{sec:gadget-construction}). The gadget measurement can however be flipped by a single measurement error on the $\chi$ qubits. To protect against these $\chi$ measurement errors, we may repeat the gadget syndrome extraction for $R$ rounds - with $R\geq d$ sufficient for measurement fault-distance matching the code distance as per Theorem 11 within Ref. \cite{Cross2025}. The aforementioned theorem concerns a phenomenological error model, without consideration of compilation to native gates or syndrome extraction scheduling. We therefore explicitly verify the circuit distance $d_c$, the minimum number of circuit errors required to result in a logical error undetected. 

For each compiled circuit, we construct its detector error model in Stim~\cite{Gidney2021Stim} and search for a minimum-cardinality set of circuit faults that produces an undetected logical failure. The search is formulated as a MaxSAT optimisation problem using Stim's \texttt{shortest\_error\_sat\_problem()} function and solved with the RC2 algorithm provided by the PySAT library~\cite{imms-sat18, itk-sat24}. The weight of the minimum fault set defines the circuit distance $d_c$ for the specified logical observable. 

For the single-gadget experiment, the circuit distance associated with the $\bar{X}_1$ gadget-measurement outcome is $d_c=1$ for $R=1$ and $d_c=3$ for $R=3$. The terminal transversal measurements of both $X$-type logical operators have circuit distance $d_c=3$. For the two-gadget experiment, the circuit distance associated with the $\bar{Z}_2$ gadget measurement outcome is $d_c=1$, whereas the terminal destructive transversal measurement of $\bar{Z}_2$ has circuit distance $d_c=3$ 

\subsection{Hardware\label{sec:hardware}}

Experiments were performed on the Quantinuum H2-2 trapped ion processor \cite{QuantinuumH2}, which provides 56 qubits with all-to-all connectivity, mid-circuit measurement and reset. We quote recent error rates of the device in \cref{tab:error-model}. Our gadget experiments utilised 27 qubits of which there were 10 data qubits, 10 ancilla qubits to perform stabiliser checks, and 7 qubits for the gadget (3 gadget qubits, 4 gadget checks), while our memory experiments required 20 qubits. Both the single- and two-gadget circuits required the same number of total qubits, as the gadget qubits (including gadget check qubits) were reused after the measurement of the first gadget. The check qubits are reset between rounds and reused. The $\kappa$ qubits are not reset for multi-round gadgets, but are reset between the X and Z gadgets in the SEQ experiment. 

Leakage errors are not corrected by the error correction protocol we adopt, and so we choose to minimise leakage at the hardware level. In all circuits we have enabled hardware leakage repump, a setting for the H2 series of processors which adds optical pumping of the leakage states back into the computational subspace during pre-gate cooling \cite{Hayes2020}. We note that having only black box access to the device means we lack certain low level compilation details: the number of pulses in the leakage repump scheme, circuit timings, ion routing and the batching of qubits into the gate and readout zones. 

\subsection{Measuring Logical Operators and Decoding}
\label{sec:logical ops decoding}

Each syndrome extraction cycle measures all of the stabilisers of the bare code, or merged code-gadget system. Initial stabiliser eigenvalues aren't necessarily deterministic, but in absence of noise they are constant between rounds. Bulk detectors are defined as the parity of stabiliser values between consecutive rounds. Boundary detectors are stabiliser parities in the first and final cycles which can't be paired across cycles, but are deterministic with respect to some other reference - the initial or final states of the data qubits. Instances when a detector reports 1, indicate the presence of an error and are labelled `detection events'. Detector layer $L$ compares cycles $L$ and $L+1$, and a circuit with $N$ syndrome extractions has $N-1$ bulk layers and $10(N-1)$ bulk detectors for the code block alone, not counting detectors in the gadget. The $\kappa_j$ outcomes enter the detectors in the trailing layer - the layer comparing the final cycle in which the gadget and code are merged and the first bare cycle after the gadget is detached. The $\kappa_j$ measurement outcomes are incorporated into the detectors of the post-gadget syndrome cycle to account for the deterministic frame shift induced by the code deformation. A visualisation for the stabiliser measurements and detectors for repeated syndrome extraction cycles is shown in Fig. \ref{fig:gadget_3_round}. 

The gadget logical observable is defined as the product of the $\chi_i$ outcomes; the reference logical observable is the product of the data qubit measurements supporting the logical representative together with any $\kappa$ outcomes required for noiseless determinism. We differentiate these observable measurements in our results using subscript notation, the logical measurement $\bar{X}_1$ inferred from the gadget is denoted as $\bar{X}_{1,G}$ and the destructive transversal measurement is labelled $\bar{X}_{1,D}$.

For both the destructive and non-destructive measurements, we require logical Pauli frame tracking to determine how the logical operators of the code are transformed during the coupling and disentangling of the gadgets. For this, we borrow the notion of dressed logical operators from subsystem codes \cite{Cohen2022, Wang2026CheckWeight, Kribs2006}. 

For the $\bar{X}_1$ gadget, the $X$-basis logical operators remain unchanged as they commute with the new stabilisers of the merged code and gadget. To understand the transformation of the $Z$-basis logical operators, we first look at their qubit support. For $\bar{Z}_2=Z_{q_2} Z_{q_6} Z_{q_7}$, data qubits $q_2$ and $q_7$ are in the support of $\chi_2$ and $\chi_7$ checks respectively (as seen in \cref{fig:gadget}). $\kappa_2$ appears in the support of $\chi_2$ and $\chi_7$ whereas $\kappa_3$ and $\kappa_4$ appear once each. The resultant fusion corrected logical representative becomes $$\bar{Z}_{2,G} \to Z_2 Z_6 Z_7\otimes Z_{\kappa_3}Z_{\kappa_4}=Z_2 Z_6 Z_7\otimes Z_{\kappa_2}$$ where the final equality is modulo $\gamma$, the gadget stabiliser. We do the same for $\bar{Z}_1=Z_{q_2} Z_{q_4} Z_{q_5}$ however there is no valid correction which commutes with all three $\chi$ checks so we choose a correction that anti-commutes with them all. Therefore, the measurement of $\bar{Z}_1$ is expected to be random. Together, the logical operators are transformed as follows:

\begin{itemize}
    \item $\bar{X}_1$ remains unchanged,
    \item $\bar{X}_2$ remains unchanged,
    \item $\bar{Z}_1 \to \bar{Z}_1 \otimes Z_{\kappa_4} \to  m(\kappa_4) \cdot \bar{Z}_1$ is maximally mixed,
    \item $\bar{Z}_2 \to \bar{Z}_2 \otimes Z_{\kappa_2} \to  m(\kappa_2) \cdot \bar{Z}_2$.
\end{itemize}

At the end of the single and three round $\bar{X}_1$ experiments, we measure the $X$-basis logical operators destructively which do not require measurement outcomes from the $\kappa$ qubits. 

For the subsequent $\bar{Z}_2$ gadget in the two-gadget experiments, we note the transformations of the logical operators are
\begin{itemize}
    \item $\bar{X}_1 \to \bar{X}_1 \otimes X_{\nu_2} \to  m(\nu_2) \cdot \bar{X}_1$,
    \item $\bar{X}_2$ is maximally mixed,
    \item $m(\kappa_4) \cdot\bar{Z}_1$ unchanged by $\bar{Z}_2$ gadget,
    \item $m(\kappa_2) \cdot \bar{Z}_2$ unchanged by $\bar{Z}_2$ gadget.
\end{itemize}

At the end of the SEQ circuit, we measure the $Z$-basis logical operators destructively. Throughout Section \ref{sec:results}, we present the number of errors for the destructive and non-destructive measurements after applying these fusion corrections. 

Decoding is performed offline with the Tesseract most-likely-error decoder \cite{Beni2025Tesseract} acting on the detector error model generated by Stim \cite{Gidney2021Stim}. In principle real-time decoding is possible on H2 using a custom built decoder, though we leave this for future work as the circuits here require no adaptive branching. Specifically, we configure the decoder to the short beam parameters as described in Ref. \cite{Beni2025Tesseract}. The decoder's predicted flip for each observable is applied to the measured value, and the disagreement rate between the corrected gadget outcome and the intended logical preparation or corrected destructive outcome defines the gadget logical measurement error. Logical error rates are bounded in equal-tailed 95\% Jeffreys credible intervals. Where we present statistics on individual detector rates we also use Jeffreys 95\% intervals, and where we average over subsets of the detectors we present 95\% bootstrap \cite{bootstrap} intervals (\SI{e4} resamples) to account for detector correlations within a shot. 

In Appendix~\ref{app:b}, we describe the methodology for generating the detector error model compiled to the Quantinuum H2-2 gate set and verifying its circuit distance. 

\section{Results\label{sec:results}}
\subsection{Single $\bar{X}_{1}$ Gadget Experiments}
\subsubsection{Gadget Operation\label{sec:gadget-operation}}

\renewcommand{\arraystretch}{1.3}
\begin{table}[]
\begin{tabular}{cccccc}
\cline{3-6}
 & \multicolumn{1}{c|}{} & \multicolumn{4}{c|}{\textbf{Prep. $\ket{++}_L$}} \\ \cline{3-6} 
 & \multicolumn{1}{c|}{} & \multicolumn{2}{c|}{\textbf{$\bar{X}_{1,D}$}} & \multicolumn{2}{c|}{\textbf{$\bar{X}_{2,D}$}} \\ \cline{3-6} 
 & \multicolumn{1}{c|}{} & \multicolumn{1}{c|}{$+$} & \multicolumn{1}{c|}{$-$} & \multicolumn{1}{c|}{$+$} & \multicolumn{1}{c|}{$-$} \\ \hline
\multicolumn{1}{|c|}{\multirow{2}{*}{\textbf{$\bar{X}_{1,G}$}}} & \multicolumn{1}{c|}{$+$} & \multicolumn{1}{c|}{481} & \multicolumn{1}{c|}{0} & \multicolumn{1}{c|}{481} & \multicolumn{1}{c|}{0} \\ \cline{2-6} 
\multicolumn{1}{|c|}{} & \multicolumn{1}{c|}{$-$} & \multicolumn{1}{c|}{6} & \multicolumn{1}{c|}{0} & \multicolumn{1}{c|}{6} & \multicolumn{1}{c|}{0} \\ \hline
\multicolumn{1}{l}{} & \multicolumn{1}{l}{} & \multicolumn{1}{l}{} & \multicolumn{1}{l}{} & \multicolumn{1}{l}{} & \multicolumn{1}{l}{} \\ \cline{3-6} 
 & \multicolumn{1}{c|}{} & \multicolumn{4}{c|}{\textbf{Prep. $\ket{--}_L$}} \\ \cline{3-6} 
 & \multicolumn{1}{c|}{} & \multicolumn{2}{c|}{\textbf{$\bar{X}_{1,D}$}} & \multicolumn{2}{c|}{\textbf{$\bar{X}_{2,D}$}} \\ \cline{3-6} 
 & \multicolumn{1}{c|}{} & \multicolumn{1}{c|}{$+$} & \multicolumn{1}{c|}{$-$} & \multicolumn{1}{c|}{$+$} & \multicolumn{1}{c|}{$-$} \\ \hline
\multicolumn{1}{|c|}{\multirow{2}{*}{\textbf{$\bar{X}_{1,G}$}}} & \multicolumn{1}{c|}{$+$} & \multicolumn{1}{c|}{0} & \multicolumn{1}{c|}{1} & \multicolumn{1}{c|}{0} & \multicolumn{1}{c|}{1} \\ \cline{2-6} 
\multicolumn{1}{|c|}{} & \multicolumn{1}{c|}{$-$} & \multicolumn{1}{c|}{0} & \multicolumn{1}{c|}{483} & \multicolumn{1}{c|}{0} & \multicolumn{1}{c|}{483} \\ \hline
\multicolumn{1}{l}{} & \multicolumn{1}{l}{} & \multicolumn{1}{l}{} & \multicolumn{1}{l}{} & \multicolumn{1}{l}{} & \multicolumn{1}{l}{} \\ \cline{3-6} 
 & \multicolumn{1}{c|}{} & \multicolumn{4}{c|}{\textbf{Prep. $\ket{00}_L$}} \\ \cline{3-6} 
 & \multicolumn{1}{c|}{} & \multicolumn{2}{c|}{\textbf{$\bar{X}_{1,D}$}} & \multicolumn{2}{c|}{\textbf{$\bar{X}_{2,D}$}} \\ \cline{3-6} 
 & \multicolumn{1}{c|}{} & \multicolumn{1}{c|}{$+$} & \multicolumn{1}{c|}{$-$} & \multicolumn{1}{c|}{$+$} & \multicolumn{1}{c|}{$-$} \\ \hline
\multicolumn{1}{|c|}{\multirow{2}{*}{\textbf{$\bar{X}_{1,G}$}}} & \multicolumn{1}{c|}{$+$} & \multicolumn{1}{c|}{119} & \multicolumn{1}{c|}{1} & \multicolumn{1}{c|}{55} & \multicolumn{1}{c|}{65} \\ \cline{2-6} 
\multicolumn{1}{|c|}{} & \multicolumn{1}{c|}{$-$} & \multicolumn{1}{c|}{4} & \multicolumn{1}{c|}{116} & \multicolumn{1}{c|}{61} & \multicolumn{1}{c|}{59} \\ \hline
\multicolumn{1}{l}{} & \multicolumn{1}{l}{} & \multicolumn{1}{l}{} & \multicolumn{1}{l}{} & \multicolumn{1}{l}{} & \multicolumn{1}{l}{} \\ \cline{3-6} 
 & \multicolumn{1}{c|}{} & \multicolumn{4}{c|}{\textbf{Prep. $\ket{11}_L$}} \\ \cline{3-6} 
 & \multicolumn{1}{c|}{} & \multicolumn{2}{c|}{\textbf{$\bar{X}_{1,D}$}} & \multicolumn{2}{c|}{\textbf{$\bar{X}_{2,D}$}} \\ \cline{3-6} 
 & \multicolumn{1}{c|}{} & \multicolumn{1}{c|}{$+$} & \multicolumn{1}{c|}{$-$} & \multicolumn{1}{c|}{$+$} & \multicolumn{1}{c|}{$-$} \\ \hline
\multicolumn{1}{|c|}{\multirow{2}{*}{\textbf{$\bar{X}_{1,G}$}}} & \multicolumn{1}{c|}{$+$} & \multicolumn{1}{c|}{122} & \multicolumn{1}{c|}{2} & \multicolumn{1}{c|}{63} & \multicolumn{1}{c|}{61} \\ \cline{2-6} 
\multicolumn{1}{|c|}{} & \multicolumn{1}{c|}{$-$} & \multicolumn{1}{c|}{1} & \multicolumn{1}{c|}{120} & \multicolumn{1}{c|}{61} & \multicolumn{1}{c|}{60} \\ \hline
\end{tabular}
\caption{Confusion matrices to illustrate correlations between gadget measurement outcome $\bar{X}_{1,G}$ and destructive measurements $\bar{X}_{1,D}$ and  $\bar{X}_{2,D}$. Results are after correction and verification. Number of shots for each preparation state are: 487 for $\ket{++}_L$, 484 for $\ket{--}_L$, 240 for $\ket{00}_L$ and 245 for $\ket{11}_L$.}
\label{tab:conf_x}
\end{table}

We first establish that a single round of the gadget measurement extracts the correct logical eigenvalue. Logical states $\ket{{0}{0}}_L$, $\ket{{+}{+}}_L$, $\ket{{-}{-}}_L$ and $\ket{{1}{1}}_L$ were prepared with preparation verified by trivial syndrome on SE1, then subjected to a single round of $\bar{X}_1$ gadget measurement, followed by a bare syndrome extraction (SE2) and destructive measurement in the $X$ basis. The verification has little impact on failure rate in all of the single gadget-round circuits, but significantly improved the three round performance which we return to in Section \ref{sec:ft}. We present the error rates of the verified runs in all cases for consistency. Verification rejection rates for all circuits are less than $5\%$ with exact rates and performance without the verification step is reported in Appendix~\ref{app:d}. 

\cref{fig:gadget-operation} shows the expectation value of $\langle \bar{X}_{1,G}\rangle$ inferred from the parity of the gadget checks $\chi_2\chi_3\chi_7$ for each preparation. For the $\bar{X}_1$ eigenstate preparations ($\ket{{+}{+}}_L$, $\ket{{-}{-}}_L$) the gadget outcome $\prod_i \chi_i$ is deterministic up to errors, and we observe an error rate of 0.72 [0.32, 1.41]\% with respect to the intended logical preparation.

\begin{figure}[h]
    \centering
    \includegraphics[width=\linewidth]{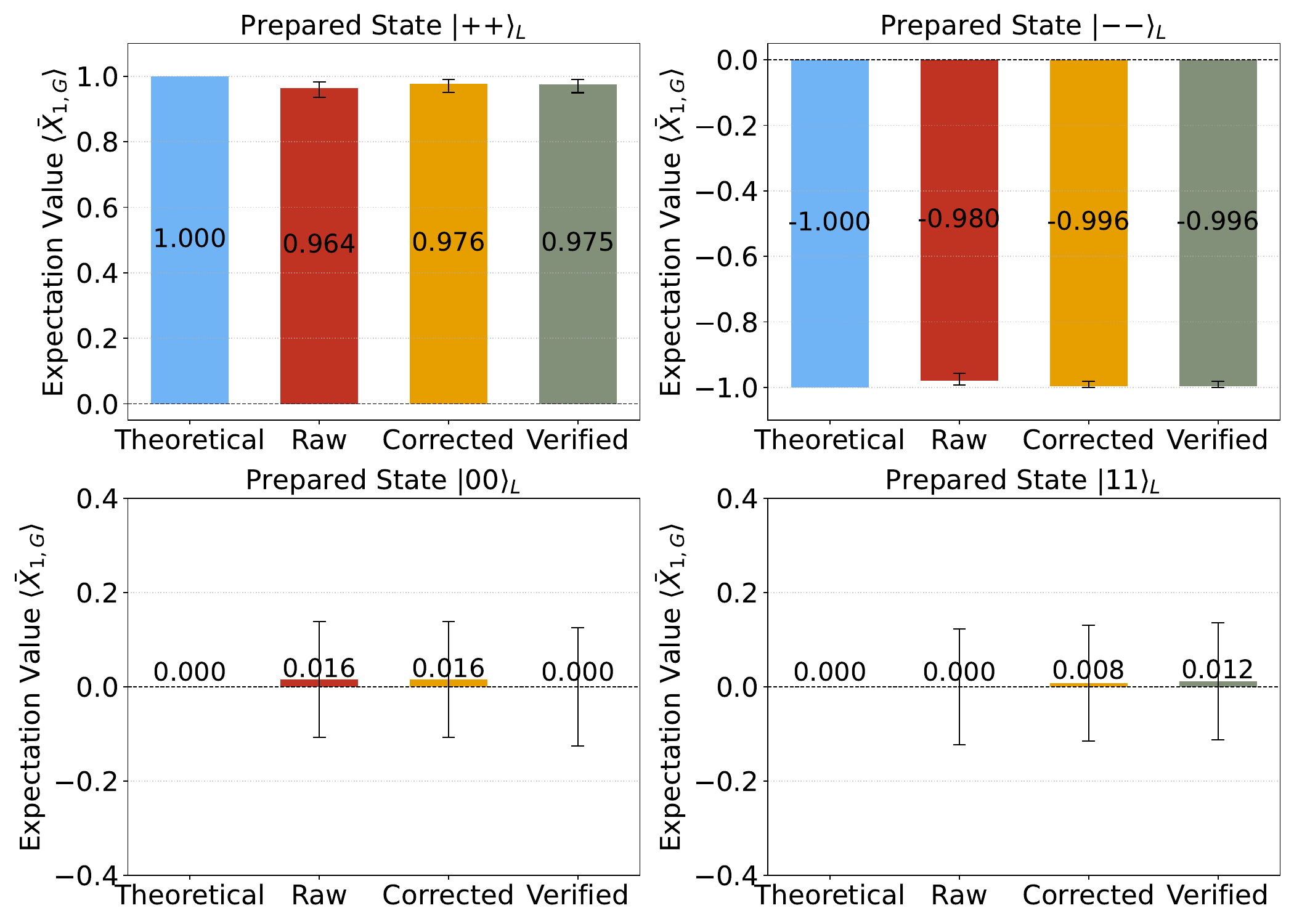}
    \caption{Expectation of  $\langle \bar{X}_{1,G}\rangle$ for different logical state preparations, compared to the ideal value (\textit{Theoretical}). \textit{Raw} is directly inferred from $\chi_2\chi_3\chi_7$, \textit{Corrected} is the Tesseract decoded result, \textit{Verified} is the \textit{Corrected} result calculated only from the shots passing verification. Error bars are 95\% Jeffreys intervals. \label{fig:gadget-operation}}
\end{figure}

\cref{tab:conf_x} displays confusion matrices of the gadget measurement against the subsequent destructive measurement outcomes for each logical preparation state. We observe that the gadget outcomes for $\ket{00}_L$, $\ket{11}_L$ are consistent with uniform randomness and that the gadget inferred random measurement in a given shot agrees with the subsequent destructive measurement in 98.4\% of shots, demonstrating projection into the state reported by the gadget, while the spectator qubit outcome is uncorrelated with the gadget outcome. Averaging over X and Z basis preparations, the gadget outcome disagrees with the subsequent destructive $X$-basis logical measurement at a rate of 1.03 [0.60, 1.65]\%.

The $\bar{X}_{1,G}$ observable in XR1 has a circuit distance of $d_c=1$ so the leading order contribution to the gadget error rate is the set of faults which flip the observable without a detection event. The probability of logical error without any detectable error can be read directly from the detector error model which we decompose into the corresponding circuit errors in \cref{tab:r1_budget}, from a physical error model based on H2-2 performance validation which we summarise in \cref{tab:error-model}. We observe that reset and measurement errors have the largest contribution, followed by 2-qubit gate errors. For each error, we count the number of independent Pauli product events that flip the logical observable and their standalone contribution to the probability of an undetectable error. The odd-parity (XOR) product of this gives a value of $6.00 \times 10^{-3}$. 

The gadget error budget in \cref{tab:r1_budget} is compatible with the observed experimental error of 0.72 [0.32, 1.41]\% and accounts for the majority of the error seen in the single round simulation in \cref{fig:sim_v_hw_erates}. These single fault contributions are suppressed by increasing the circuit distance from 1 to 3 by decoding over 3 rounds of syndrome measurement.

\begin{table}[h]
    \centering
        \renewcommand{\arraystretch}{1.5} 
    \begin{tabular}{|c|c|c|c|}
        \hline
         \textbf{Error} & \textbf{Probability} & \textbf{Events} & \textbf{Contribution}\\
         & $(\times 10^{-4})$ & & $(\times 10^{-3})$\\
        \hline
        Reset \& Measure & $4.70$ & $6$ & $2.81$ \\        
        Idle (SE1) & $1.89$ & $3$ & $0.57$ \\
        Idle (Code+Gadget) & $2.57$ & 3 & $0.77$ \\
        1Q Gate & $0.11$ & $24$ & $0.27$ \\
        2Q Gate & $0.67$ & $24$ & $1.60$ \\
        \hline
        Total Contribution & & & $6.00$ \\
        \hline
    \end{tabular}
    \caption{Error budget decomposition of independent Pauli events and their contribution to the probability of an undetectable logical error. Total derived from odd-parity product not sum of table rows. }
    \label{tab:r1_budget}
\end{table}

\subsubsection{Fault-Tolerance - Repeated Gadget Measurement}
\label{sec:ft}

\begin{figure*}
\centering
    \centering
    \begin{subfigure}[t]{0.66\textwidth}
        \centering
        \includegraphics[width=\linewidth]{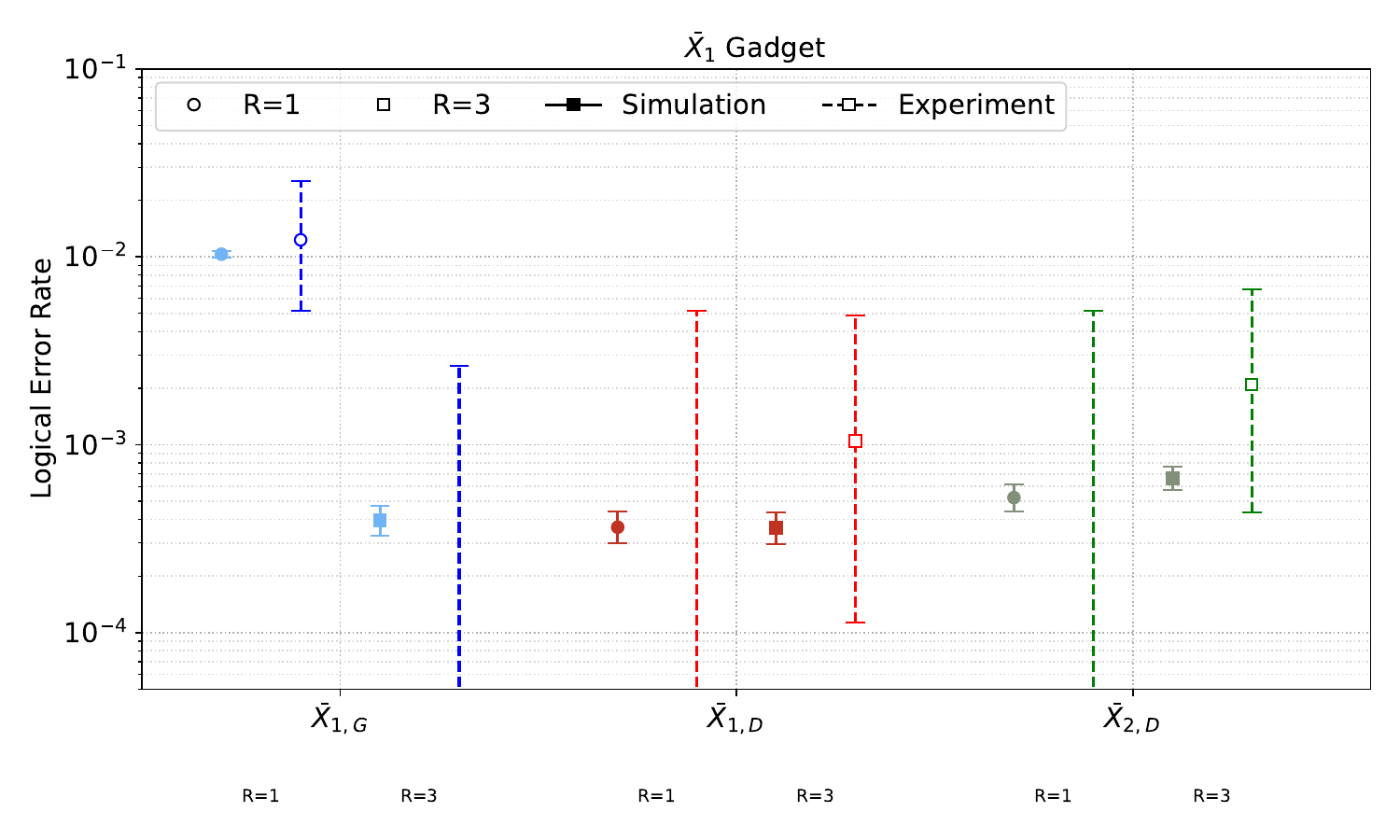}
        \caption{}
    \label{fig:sim_v_hw_erates}
    \end{subfigure}%
    ~ 
    \begin{subfigure}[t]{0.33\textwidth}
        \centering
        \includegraphics[width=\linewidth]{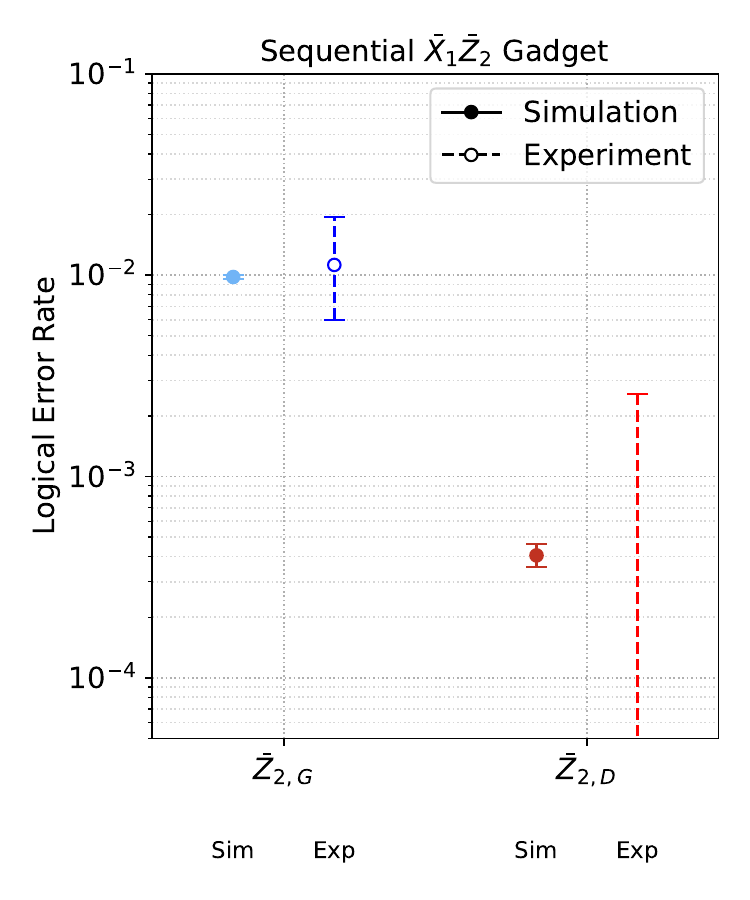}
        \caption{}
    \end{subfigure}
    \caption{(a) Measurement results of $\bar{X}_1$ from single gadget circuit with $\ket{++}_L$ logical data preparation. Experimental results are presented post-verification with 487 shots for $R=1$ and 957 shots for $R=3$. (b) Measurement results of $\bar{Z}_2$ from sequential $\bar{X}_1\bar{Z}_2$ gadget circuit with $\ket{00}_L$ and $\ket{11}_L$ logical data preparation. Experimental results are presented post-verification with 978 shots in total. Simulation results assume our H2-2 inspired error model with $10^6$ shots for each prepared state. Experimental results with no reported logical errors are presented with only upper bounds.}
\end{figure*}

A single measurement of the gadget checks is vulnerable to measurement faults: a single faulty $\chi_i$ outcome flips the inferred logical eigenvalue. 
To demonstrate the fault-tolerant implementation of the gadget we performed the $\bar{X}_1$ measurement over three consecutive syndrome extraction rounds, for the $\ket{++}_L$ $X$-basis preparation, decoding the entire 3 round 62 detector syndrome (full syndrome shown in Appendix~\ref{app:a}). We emphasise this is not a majority voting on the repeated rounds, which would treat the repetitions as an independent classical bit without utilising the inter-round correlations on the gadget stabilisers to define detectors. In the repeated gadget measurements, detectors are defined on the  $\chi_i$ and $\gamma$ checks between rounds, a single faulty $\chi_i$ readout is then corrected by the decoder.

Figure \ref{fig:sim_v_hw_erates} compares the logical measurement error of the single gadget-round and three gadget-round protocols. For the XR3 gadget circuit, only $\ket{++}_L$ was prepared, with 0 errors in 957 verification passing shots consistent with simulation and upper bounding the error rate at 0.26\% in a 95\% credible interval. For the single round circuit, considering the $\ket{++}_L$ preparation shots alone there are 6 errors in 487 verified shots and the error rate is 1.23 [0.52, 2.52]\%. At the reported shot counts the simulated error rate for the XR3 gadget experiment can not be resolved, however the three round error rate is bounded below the credible interval of the single round, demonstrating a reduction in error rate. The presence of the weight-1 failure modes of \cref{tab:r1_budget} alone would result in an error rate \num{6.00e-3}, and an expected 6 errors in 957 trials (probability of recording 0 decoding failures as observed $\sim 0.3$\%). To separate the effect of repeated measurement of the gadget from variation between the XR1 and XR3 results series, we repeat decoding of the $R=3$ data omitting the 6 detectors which compare $\chi$ measurements between rounds (Full detector set shown in Appendix A Fig. \ref{fig:gadget_3_round}), removing the ability to detect measurement errors on $\chi$. Decoding without the $\chi$ detectors results in 7/957 errors (error rate 0.73 [0.33, 1.43]\%) consistent with the presence of the weight-1 failure modes and $R=1$ error rate, validating the attribution of the $R=3$ error rate to suppression of the measurement fault channels. 

\begin{figure*}[t!]
\centering
\includegraphics[width=\textwidth]{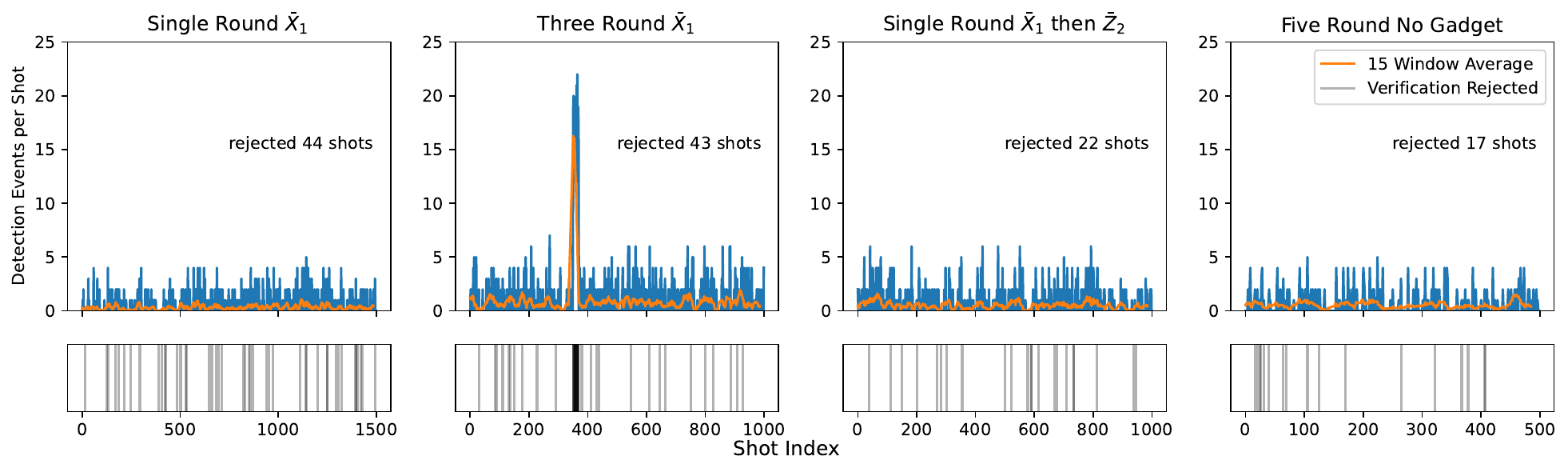}
\caption{Sum over all detection events in each shot, on the (20, 40, 30, 40) code block bulk detectors for the single and three-round $\bar{X}_{1,G}$, sequential $\bar{X}_{1,G}$ then $\bar{Z}_{2,G}$, and no-gadget memory experiments respectively. For the three round experiment, within the shots 351-369 all detectors are randomised. Verification on the first round of stabilisers filters 43 shots total and all shots 351-369. \label{fig:shotquality} }
\end{figure*}

We note that the XR3 error rate depends entirely on the verification step, without verification there are 10 errors in 1000 shots. The XR1 logical error results show no strong dependence on verification despite comparable rejection fraction 2.9\% for XR1 and 4.3\% for XR3.  \cref{fig:shotquality} shows the total bulk detection events on the code block per shot in each circuit. At the circuit level all shots should be independent however we identify a window of consecutive raised detection events. Repeating simulation of the three round circuit under the maximum error rates quoted in the H2-2 datasheet (\cite{quantinuum_h2_datasheet}, distinct from the error model rates in \cref{tab:error-model} which use typical values from \cite{quantinuum_performance_validation}), we define a threshold on total detection events at the 99th percentile of a maximum 15-shot window mean (\SI{e4} samples of 1000 shots). For the hardware we flag all shots in the centre of a 15-shot window with mean exceeding this threshold. Flagging in this way identifies a single window of 28 consecutive shots from shot count 346 to 373. 

The SE1 verification is intended to check proper logical state preparation, defined in advance and applied to all circuits. Verification on SE1 rejects 43 shots total, 19 of which are the consecutive shots 351--369 entirely within the flagged window and all of the $\bar{X}_{1,G}$ decoding failures fall within the range 351--369 i.e. within this range the $\bar{X}_{1,G}$ observable is a coin flip. Terminal transversal measurements $\bar{X}_{1,D}$ and $\bar{X}_{2,D}$ suffer similarly high error rates in the verification rejected portion of the flagged window. The verification rejects $\sim 68$\% of the flagged region vs $\sim 2.5$\% outside the flagged region. Within the shots 351--369 all detectors are randomised, indicating a loss of coherent operation.
By using a greedy scheduling of the $4$ gate zones for the $R=3$ circuit we estimate a single shot to require $\sim$ \SI{500}{ms} and the excursion in performance to represent $\sim$ \SI{10}{s}. As the event identified in the three round series spans multiple shots and affects the entire register we present the verified shots throughout to isolate the performance of the gadget. 

In summary, we have shown the single-round gadget correctly extracts the targeted eigenvalue with error rate 0.72 [0.32, 1.41]\% consistent with the the dominant error channel being weight-1 errors when reading out the $\chi$ qubits. The gadget measurement is eigenvalue preserving, agreeing with the subsequent transversal measurements at error rate 1.03 [0.60, 1.65]\%. Finally, we demonstrate the suppression of measurement error by decoding over repeated rounds of gadget measurement reporting a three-round gadget error rate $<$0.26\% (95\% CrI). 

\subsection{Code Block Preservation and Selective Addressing \label{sec:qnd}}

\begin{figure*}[t!]
\centering
\includegraphics[width=\linewidth]{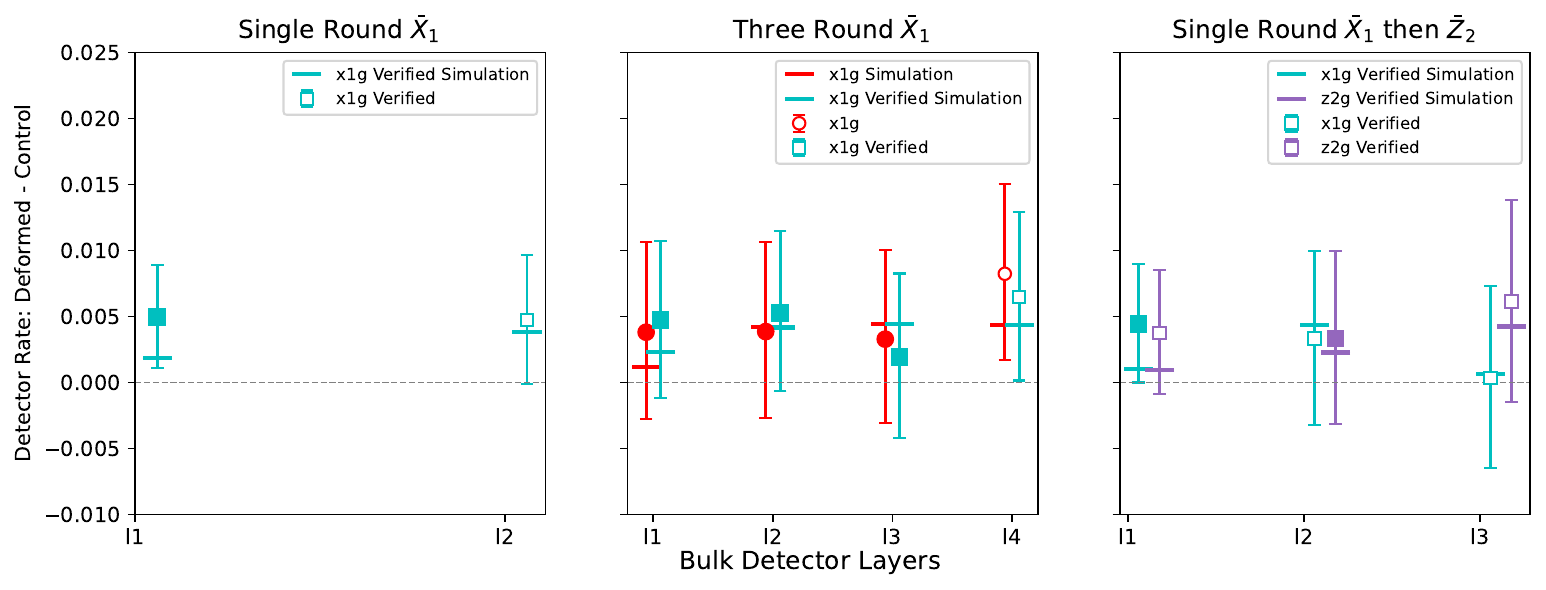}
\caption{Excess detector rate of deformed stabilisers over the control stabilisers, per detector layer, for the single round $\bar{X}_{1,G}$, three round $\bar{X}_{1,G}$ and single round sequential $\bar{X}_{1,G}$ then $\bar{Z}_{2,G}$ experiments. Filled markers show layers in which a gadget is attached, open markers show layers the corresponding gadget is detached. Square markers show experimental data on shots passing verification on trivial SE1, with 95\% bootstrap intervals (\SI{1e4} resamples for each prepared state). Excess without applying verification plotted for the three-round circuit in red with circle markers - the spike in hardware error rate identified in Fig. \ref{fig:shotquality} affects the deformed and control group proportionately.  Horizontal line markers show simulation at rates in \cref{tab:error-model}. Simulation verification is as described for hardware. The dashed line shows zero excess.  } \label{fig:excess}
\end{figure*}

A defining feature of the surgery measurement is that it is non-destructive to the code block, and logical information not addressed by the gadget is preserved, in contrast with transversal measurement. We assess this in two ways: through the syndrome during and after merge and through a subsequent gadget measurement on the spectator qubit. 

\subsubsection{Syndrome Analysis}
Bulk detector layers compare stabiliser measurements from two subsequent syndrome extraction cycles, layer $L$ compares cycles $L$ and $L+1$. In the absence of noise, each syndrome extraction cycle is deterministic and every detector comparing consecutive extraction rounds returns 0. To assess whether the system remains in the code space, we analyse the detector rates, the fraction of shots in which the detector returns 1, comparing the results from the gadget experiments against the MEM circuit and against simulation at the error rates in \cref{tab:error-model}. We partition the set of code stabilisers, 
 $$S=\{S_{Z,i},S_{X,i}: i\in\{0,1,2,3,4\}\}$$
of the \code{10}{2}{3} code into 3 sets, the Z checks deformed by the $\bar{X}_1$ gadget, $$\text{x1g}=\{S_{Z,2},S_{Z,3},S_{Z,4}\},$$ the $\chi$ checks deformed by the $\bar{Z}_2$ gadget, $$\text{z2g}=\{S_{X,0},S_{X,1},S_{X,2}\}$$ and a control set. For each circuit we define the control set $C$ as the stabilisers not deformed by any gadget in that circuit: for circuits with no gadget (MEM), $C=S$, for the XR1 and XR3 circuits control is defined as $C=S\setminus \text{x1g}$, and for the SEQ circuit control is $C=S\setminus \left(\text{x1g} \cup \text{z2g}\right)$.

\begin{table}
    \centering
    \renewcommand{\arraystretch}{1.5} 
    \begin{tabular}{|c|c|c|c|c|}
        \hline
        \textbf{Circuit} & \textbf{Set}  & \textbf{Layer} &  \textbf{No Fusion [\%]}  &  \textbf{Fusion [\%]}   \\
        \hline
        XR1&x1g& L2 &  49.6 [47.0, 52.2]  & 2.06 [1.42, 2.89]   \\
        
        XR3&x1g& L4 &  47.1 [44.0, 50.3]  & 2.51 [1.66, 3.65]   \\
       
        SEQ&x1g& L2 & 50.1 [47.0, 53.2]  & 1.94 [1.21, 2.96]  \\
       
        SEQ&z2g& L3 &  48.1 [44.9, 51.2]  & 2.66 [1.79, 3.81]  \\
        \hline
    \end{tabular}
    \caption{Trailing layer detector rates on verified shots, for the set of deformed stabilisers. Fusion correction recovers randomised detectors indicating return to \code{10}{2}{3} code space. Values are minimum over set without fusion correction (No Fusion) and maximum over set with fusion correction (Fusion). Jeffreys 95\% intervals.}
    \label{tab:fusion}
\end{table}

To see that the code is returned to the \code{10}{2}{3} code space after the gadget is detached we inspect the detector rate on deformed stabilisers in the trailing layer, the layer comparing the final gadget syndrome to the first syndrome after the gadget is detached. The trailing layer contains the fusion corrected stabilisers, where the $X$-type gadget's $\kappa$ qubits ($Z$-type gadget's $\nu$ qubits) measurement outcomes are included in the x1g (z2g) stabiliser parities.  Table \ref{tab:fusion} shows the maximum detector rate over the deformed stabiliser sets for each circuit against the minimum  rate across the same set in absence of the fusion correction. Without the fusion correction we see the detectors are randomised, whereas with the appropriate fusion correction the detectors are made deterministic up to circuit noise. The maximum detector rates are (2.06, 2.51, 1.94, 2.66)\% for the single-round, three-round and two-gadget circuit's x1g stabilisers and the two-gadget circuit's z2g stabilisers respectively. If the system leaves the code space throughout the gadget circuit this would also be seen by increased detector rate. \cref{tab:max_rates} shows the single highest rate detector across all layers in each circuit, its associated rate, as well as the total number of detectors. The maximum single detector rate for gadget circuits is 3.80\% (2.82\%) against 2.80\% (2.90\%) for the MEM circuit across all shots (across verified shots). The only circuit for which a single detector's rate is appreciably elevated above the maximum rate in the MEM circuit is XR3 circuit without verification - attributed to the spike in machine error seen in \cref{fig:shotquality}. 

\begin{table}
    \centering
    \renewcommand{\arraystretch}{1.5}
    \begin{tabular}{|c|c|c|c|c|}
        \hline
        \textbf{Circuit} & \textbf{Shots} & \textbf{Detector} & \textbf{Rate [\%]} & \textbf{\# Bulk} \\
        \hline
        XR1 & all & (L1, $S_{Z,4}$) & 2.27 [1.60, 3.11]  & 20 \\
        XR1 & verified & (L2, $S_{Z,4}$) & 2.06 [1.42, 2.89] & 20 \\
        XR3 & all & (L4, $S_{Z,2}$) & 3.80 [2.75, 5.12] & 40 \\
        XR3 & verified & (L3, $S_{Z,2}$) & 2.82 [1.91, 4.02] & 40 \\
        SEQ & all & (L3, $S_{X,0}$) & 2.70 [1.83, 3.85] & 30 \\
        SEQ& verified & (L3, $S_{X,0}$) & 2.66 [1.79, 3.81] & 30 \\
        MEM & all & (L2, $S_{X,3}$) & 2.80 [1.61, 4.53]& 40 \\
        MEM & verified & (L2, $S_{X,3}$) & 2.90 [1.67, 4.69] & 40 \\
        \hline
    \end{tabular}
    \caption{Maximum single detector rate over all circuits. The detector column provides circuit location as (layer, stabiliser), for each circuit. Maximum is over the bulk code detectors in each circuit, enumerated in the final column (\# Bulk). Jeffreys 95\% intervals. }
    \label{tab:max_rates}
\end{table}

 The control group rates agree with simulation at the rates in \cref{tab:error-model} on average (Appendix \ref{app:c}, Fig \ref{fig:control_group}), but vary between layers beyond the sampling error, including in the no-gadget memory circuit. This structure in detector rate across layers is not captured by our model, which includes depth-weighted idle in each stabiliser round. We attribute the difference to low-level compilation details not exposed to the user: transport schedules, circuit timings and batching of operations. 

 To separate the impact of the gadget on the deformed stabilisers from the circuit specific effects seen in the control stabilisers, we define the excess of the deformed stabiliser detectors over the control group detectors: $$E_L(\text{x1g})=r_L(\text{x1g})-r_L(C)$$ and $$E_L(\text{z2g})=r_L(\text{z2g})-r_L(C)$$ where $r_L(A)$ is the detector rate in layer $L$, averaged over detectors in set A for A $\in \{\text{x1g},\text{z2g},\text{C}\}$. \cref{fig:excess} compares $E_L$ against simulation. Here the error bars are bootstrapped 95\% intervals resampled within each state preparation, as detectors within a shot are not independent. Each circuit type is simulated over the same selection of state preparations as measured on hardware (summarised in Appendix \ref{app:d}), at 100,000 shots per preparation. Simulation shots are verified on trivial SE1 as for the experimental data. The maximum additive excess across verified shots in all circuits is 0.0064 [0.00014,0.0129]  (95\% bootstrapped interval) in layer 4 of the three-round circuit. The excess is consistent with simulation in all layers. We plot the excess without verification for the three-round data showing that the increased detector rate in this series is common to the deformed and control stabilisers and largely cancels. For the single layer in which we measure the detector rate after the trailing layer, x1g layer 3, we see the detector rate drop as simulated, to be expected if the gadget's effect on the syndrome is transient. 

In summary the gadgets cause a small excess in detector rate on the deformed stabilisers consistent with simulation. Measured detector rates remain low across all detectors in the gadget circuits, less than $2.82\%$ when rejecting unverified shots or less than $ 3.8\%$ counting all shots, across all layers indicating continued presence in the code space.  

\subsubsection{Spectator and Second Gadget}
The syndrome analysis of the previous section demonstrates the \code{10}{2}{3} block remains in the code space. The preservation of information on the spectator logical qubit, which is idle during the first gadget measurement, can be judged from the terminal transversal measurement $\bar{X}_{2,D}$. In the $\bar{X}_{1,G}$ circuits, for state preparations $\ket{++}_L$ and $\ket{--}_L$ $\bar{X}_{2,D}$ agrees with the prepared state for all 971 verified shots of the single round circuit ($<$0.26\% 95\% CrI) and disagrees in 2/957 verified shots for the three round circuit (error rate 0.21 [0.04, 0.67]\%), consistent with simulation as shown in \cref{fig:sim_v_hw_erates}. For state preparations $\ket{00}_L$ and $\ket{11}_L$ the $\bar{X}_{2,D}$ is uncorrelated with the random outcome of the $\bar{X}_{1,G}$ measurement as seen in \cref{tab:conf_x}. 

To test preservation when the spectator is in the conjugate basis to the gadget measurement, we prepare both qubits in the $Z$ basis, measure $\bar{X}_{1,G}$ using the gadget from section \ref{sec:gadget-construction} and then measure the spectator qubit with another gadget $\bar{Z}_{2,G}$, the circuit is shown at logical level in \cref{fig:X1-Z2-circuit}. The seven gadget qubits (data and check) are reused between gadget measurements, with the $\kappa$ qubits from the first gadget being reset and prepared in $\ket{+}$ as $\nu$ for the $Z$-type gadget. As $\bar{X}_1$ and $\bar{Z}_2$ commute, $\bar{Z}_2$ is deterministic in absence of noise.

Sequential gadget measurements on logical qubits with overlapping representatives (qubits 2 and 7 in this case) are not independent. Although the $\kappa$ qubits are reset between gadgets, the data qubits they are entangled with are not. Making the second gadget deterministic requires propagating a fusion correction from the first gadget. In the single-gadget experiments, the terminal measurements matched the basis of the gadget and as a result the fusion correction was only seen in the stabilisers. With two gadgets of conjugate bases, the fusion correction may propagate to the logical representatives as shown in section \ref{sec:logical ops decoding}. A correction is applied where there is overlap between data qubits in one gadget and the $\chi$ checks in the previous gadget, each $\kappa$ appearing in the support of an odd number of the $\chi$ checks in the overlap has its fusion measurement result added to the correction.

\begin{figure}[t!]
\centering
\includegraphics[width=\linewidth]{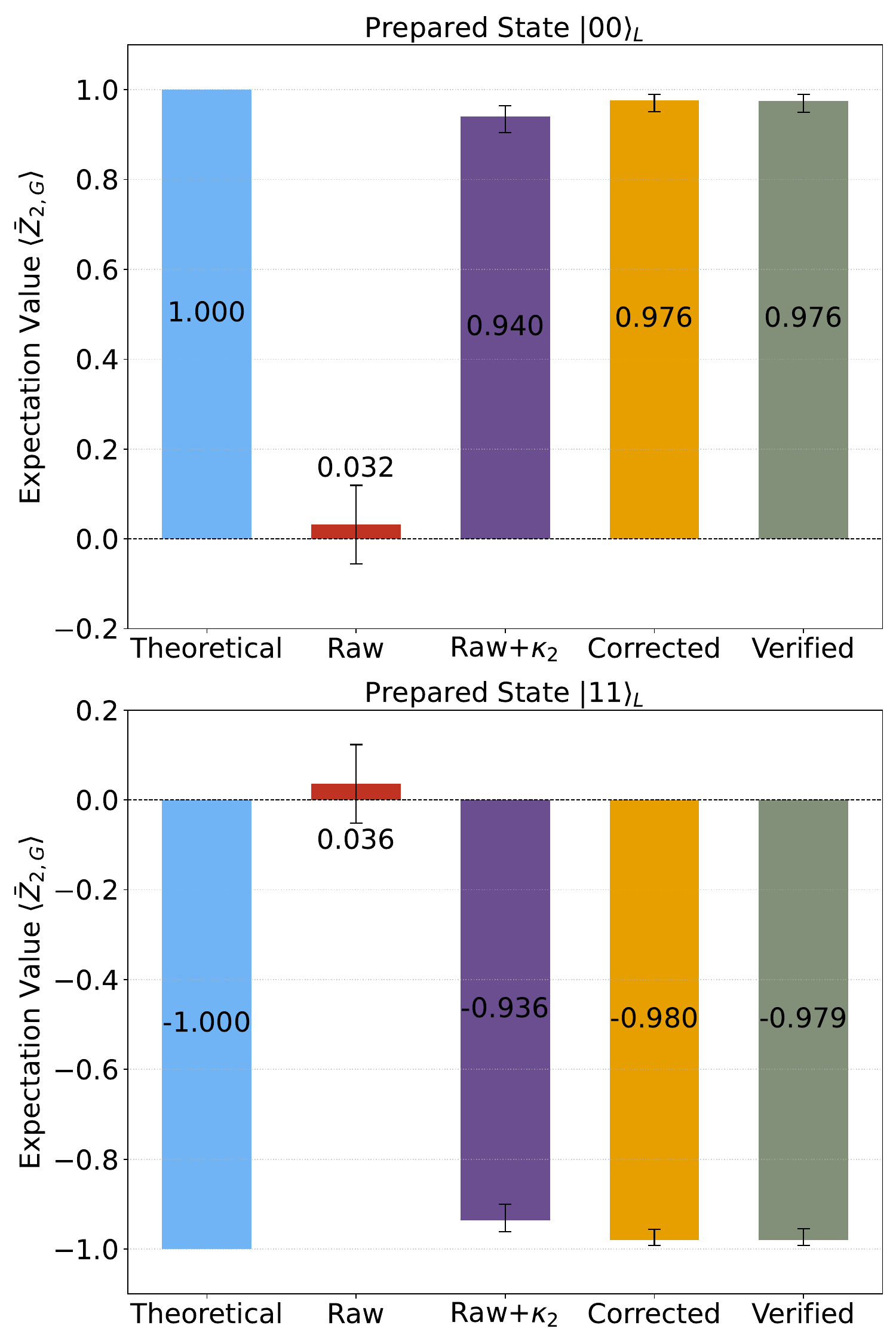}
\caption{Expectation of  $\langle \bar{Z}_{2,G}\rangle$ in the sequential gadgets circuit for different logical state preparations, compared to the ideal value (\textit{Theoretical}). \textit{Raw} is directly inferred from $\eta_2\eta_6\eta_7$, \textit{Raw+$\kappa_2$} shows the fusion-corrected result incorporating the $\kappa_2$ measurement from the previous gadget, \textit{Corrected} is the fusion-corrected result decoded with Tesseract, \textit{Verified} is the \textit{Corrected} result calculated only from the shots passing verification. The raw result is close to zero, requiring the fusion correction to make $\langle \bar{Z}_{2,G}\rangle$ deterministic. Error bars are 95\% Jeffreys intervals. } \label{fig:z_gadget}
\end{figure}

\cref{fig:z_gadget} shows the expectation of $\bar{Z}_2$ inferred from a second single round ($d_c=1$) gadget measurement on the spectator logical qubit from the first gadget measurement.  The necessity of the fusion correction can be seen by the nominally zero expectation value for the raw $\bar{Z}_2$ gadget parity before adding the $\kappa$ correction - the next column shows the expectation given the appropriate fusion correction but without application of a decoder for QEC. The final columns show the $\kappa$ corrected, decoded expectation with and without a verification step rejecting shots with non-trivial SE1. Averaging over preparation in $\ket{00}_L$ and $\ket{11}_L$ the $\bar{Z}_2$ gadget measurement disagrees with preparation in 11/978 verified shots, with error rate 1.12 [0.60, 1.94]\%.

\cref{tab:conf_z} reports the joint outcomes of the $\bar{Z}_2$ gadget and the destructive $Z$-basis readout. The destructive $\bar{Z}_{2,D}$ agrees with the intended preparation in all 978 verified shots limiting the probability the spectator's eigenvalue is flipped by the X-gadget, fusion correction or Z-gadget to $<$ 0.26\%. The instances of $\bar{Z}_{2,G}$ disagreeing with preparation are then attributed to the weight-1 measurement errors on the gadget's $\eta$ checks, equivalent to the $\chi$ measurement errors in table \ref{tab:r1_budget}.

Simulation at the parameters in \cref{tab:error-model} supports the attribution to measurement error. We simulate \SI{e6} shots preparing $\ket{00}_L$ and \SI{e6} preparing $\ket{11}_L$ before measuring the $\bar{Z}_{2,G}$ gadget at the same circuit location (cf. \cref{fig:X1-Z2-circuit}) as in SEQ with logical error rate 1.000 [0.986, 1.014]\%. The simulated values are consistent with the experimentally measured $\bar{Z}_{2,G}$ error rate 1.12 [0.60, 1.94]\%. The $\bar{Z}_{2,G}$ values are consistent with the single-round $\bar{X}_{1,G}$ logical error rate 0.72 [0.32, 1.41]\%, expected as the Z gadget carries the same leading order contributions to the X gadget (see \cref{tab:r1_budget}), up to single qubit rotations entering at the \SI{e-4} level and a shorter idle time (the first gadget's qubits have an SE1 idle contribution but the second gadget's qubits are initialised immediately before use). 

Finally, the $\bar{Z}_{1,D}$ results are consistent with uniform randomness due to the previous $\bar{X}_{1,G}$ measurement. Taken together with the determinism of $\bar{Z}_{2,G}$ this establishes the individual addressing of the surgery protocol. Information conjugate to the gadget's targeted operator collapses, whilst the commuting logical information remains addressable, for logical qubits sharing a code block, overlapping on two qubits and measured with the same gadget qubits. 

Taken together the results of this section demonstrate the individual addressing and code space preserving properties of the gadget measurements, and specifically that this allows for composition as demonstrated. Sequential logical measurement is required for surgery-based Pauli computing where one logical measurement is performed per logical cycle \cite{Webster2026Pinnacle,Yoder2025TourDeGross,He2025Extractors}. \cref{fig:X1-Z2-circuit} is a minimal example of this composition. 

\renewcommand{\arraystretch}{1.3}
\begin{table}[]
    \begin{tabular}{cccccc} \cline{3-6} & \multicolumn{1}{c|}{} & \multicolumn{4}{c|}{\textbf{Prep. $\ket{00}_L$}} \\ \cline{3-6} & \multicolumn{1}{c|}{} & \multicolumn{2}{c|}{\textbf{$\bar{Z}_{1,D}$}} & \multicolumn{2}{c|}{\textbf{$\bar{Z}_{2,D}$}} \\ \cline{3-6} & \multicolumn{1}{c|}{} & \multicolumn{1}{c|}{0} & \multicolumn{1}{c|}{1} & \multicolumn{1}{c|}{0} & \multicolumn{1}{c|}{1} \\ \hline \multicolumn{1}{|c|}{\multirow{2}{*}{\textbf{$\bar{Z}_{2,G}$}}} & \multicolumn{1}{c|}{0} & \multicolumn{1}{c|}{252} & \multicolumn{1}{c|}{234} & \multicolumn{1}{c|}{486} & \multicolumn{1}{c|}{0} \\ \cline{2-6} \multicolumn{1}{|c|}{} & \multicolumn{1}{c|}{1} & \multicolumn{1}{c|}{3} & \multicolumn{1}{c|}{3} & \multicolumn{1}{c|}{6} & \multicolumn{1}{c|}{0} \\ \hline \multicolumn{1}{l}{} & \multicolumn{1}{l}{} & \multicolumn{1}{l}{} & \multicolumn{1}{l}{} & \multicolumn{1}{l}{} & \multicolumn{1}{l}{} \\ \cline{3-6} & \multicolumn{1}{c|}{} & \multicolumn{4}{c|}{\textbf{Prep. $\ket{11}_L$}} \\ \cline{3-6} & \multicolumn{1}{c|}{} & \multicolumn{2}{c|}{\textbf{$\bar{Z}_{1,D}$}} & \multicolumn{2}{c|}{\textbf{$\bar{Z}_{2,D}$}} \\ \cline{3-6} & \multicolumn{1}{c|}{} & \multicolumn{1}{c|}{0} & \multicolumn{1}{c|}{1} & \multicolumn{1}{c|}{0} & \multicolumn{1}{c|}{1} \\ \hline \multicolumn{1}{|c|}{\multirow{2}{*}{\textbf{$\bar{Z}_{2,G}$}}} & \multicolumn{1}{c|}{0} & \multicolumn{1}{c|}{2} & \multicolumn{1}{c|}{3} & \multicolumn{1}{c|}{0} & \multicolumn{1}{c|}{5} \\ \cline{2-6} \multicolumn{1}{|c|}{} & \multicolumn{1}{c|}{1} & \multicolumn{1}{c|}{244} & \multicolumn{1}{c|}{237} & \multicolumn{1}{c|}{0} & \multicolumn{1}{c|}{481} \\ \hline 
    \end{tabular}
     \caption{Confusion matrices to illustrate correlations between gadget measurement outcome $\bar{Z}_{2,G}$ and destructive measurements $\bar{Z}_{1,D}$ and  $\bar{Z}_{2,D}$. Results are after correction and verification.}
    \label{tab:conf_z}
\end{table}

\section{Conclusion and Outlook}
We have demonstrated individually addressed, non-destructive logical Pauli measurement via QLDPC surgery on a trapped ion processor. We use a 7 qubit gadget to measure $\bar{X}_1$ on a \code{10}{2}{3} generalised bicycle code. A single round of gadget syndrome extraction returns the correct eigenvalue with an error of 0.72 [0.32, 1.41]\% relative to the verified logical preparation, and 1.03 [0.60, 1.65]\% relative to a subsequent destructive readout, in agreement with simulation under a circuit error model based on device performance validation data from Quantinuum. 

The detector rate for gadget deformed stabilisers in the layer trailing gadget detachment is less than 2.66 [1.79, 3.81]\% across circuits against $\sim50$\% without the fusion correction, indicating proper frame tracking and return to the \code{10}{2}{3} code space. The additive excess in detector rate of the deformed stabilisers over control is less than 0.0064 [0.00014, 0.0129] across verified shots and consistent with simulation. The detector rates remain low throughout the gadget circuits with a single detector maximum of 2.82 [1.91, 4.02]\% compared with a no-gadget baseline of 2.90 [1.67, 4.69]\% across verified shots. 

Measurement eigenvalues on the spectator are preserved with 0/971, 2/957, 0/978 logical errors in the transversal readout of the second logical qubit for the XR1, XR3 and SEQ circuits respectively. A second gadget applied to the spectator logical qubit recovers $\bar{Z}_2$ at 1.12 [0.60, 1.94]\%. Repeating the syndrome extraction over $R=3$ rounds and decoding the full 62 detector's history raises the circuit distance of the logical measurement to $d_c=3$. We demonstrate 3 round decoding for preparation in state $\ket{++}_L$ measuring 0 errors in 957 verified shots, bounding the logical error rate below 0.26\%, excluding the single weight errors of the $R=1$ implementation. We also decode the same XR3 data with the detectors comparing $\chi$ between rounds removed resulting in 7/957 errors consistent with the XR1 error rate. We therefore attribute the improvement in error rate of XR3 to fault-tolerance to $\chi$ readout errors. 

We have demonstrated the generalised surgery primitive at minimal distance limiting error correction performance, and the decoding is offline. There are natural next steps toward Pauli-based computing on QLDPC codes: real-time decoding with feed-forward allowing for adaptivity in measurement bases; measuring Pauli-products through bridging gadgets \cite{Cross2025}; and extension to higher distance codes. Recent work from Berthusen et al. reported a $d=6$ logical memory at error rate \SI{4.6e-5} on Quantinuum's Helios processor by concatenating the \code{10}{2}{3} rotated toric code with the \code{4}{2}{2} code, in addition to demonstrating the logical Clifford group through code automorphisms and fold-transversal gates \cite{berthusen2026experimentalvalidationcompactfaulttolerant}. There is also room to improve on the time overhead incurred in suppressing measurement errors in surgery schemes through amortised constant time surgery \cite{cowtan2025fastfaulttolerantlogicalmeasurements} applied to 2D hypergraph product codes, including the toric code, in Ref. \cite{chang2026constanttimesurgery2dhypergraph}. Baspin et al. show an alternative approach to realise fast surgery for individual operations as opposed to amortised constant time, at the cost of a larger ancilla system \cite{baspin2025fastsurgeryquantumldpc}. Proposals for low-overhead architectures motivate the study of Pauli-computing on QLDPC codes. The basis of Pauli-based computing is composing Pauli measurements and the sequential two-gadget circuit reported here is the smallest instance of this composition. 

\begin{acknowledgments}
The authors thank Dan Browne for insightful suggestions and guidance and Cameron Deans, Jonathan Burnett and Gerard Milburn for helpful comments on the manuscript. We are grateful to Laura Martin and the QCaaS team for providing access to Quantinuum H2-2 through the SparQ Access programme and for technical support.
This work was supported by the UK National Quantum Computing Centre [NQCC200921] and Engineering and Physical Sciences Research Council [grant number EP/S021582/1]. Access to quantum compute was provided by the UK National Quantum Computing Centre [NQCC200921].
\end{acknowledgments}

\subsection*{Data Availability Statement}
The data that support the findings of this study are openly available at the following URL/DOI: https://doi.org/10.5281/zenodo.23165109 \cite{ExperimentalData2026}. 

\appendix

\section{Full Gadget Syndrome Processing\label{app:a}}

\begin{figure*}
    \centering
    \includegraphics[width=\linewidth]{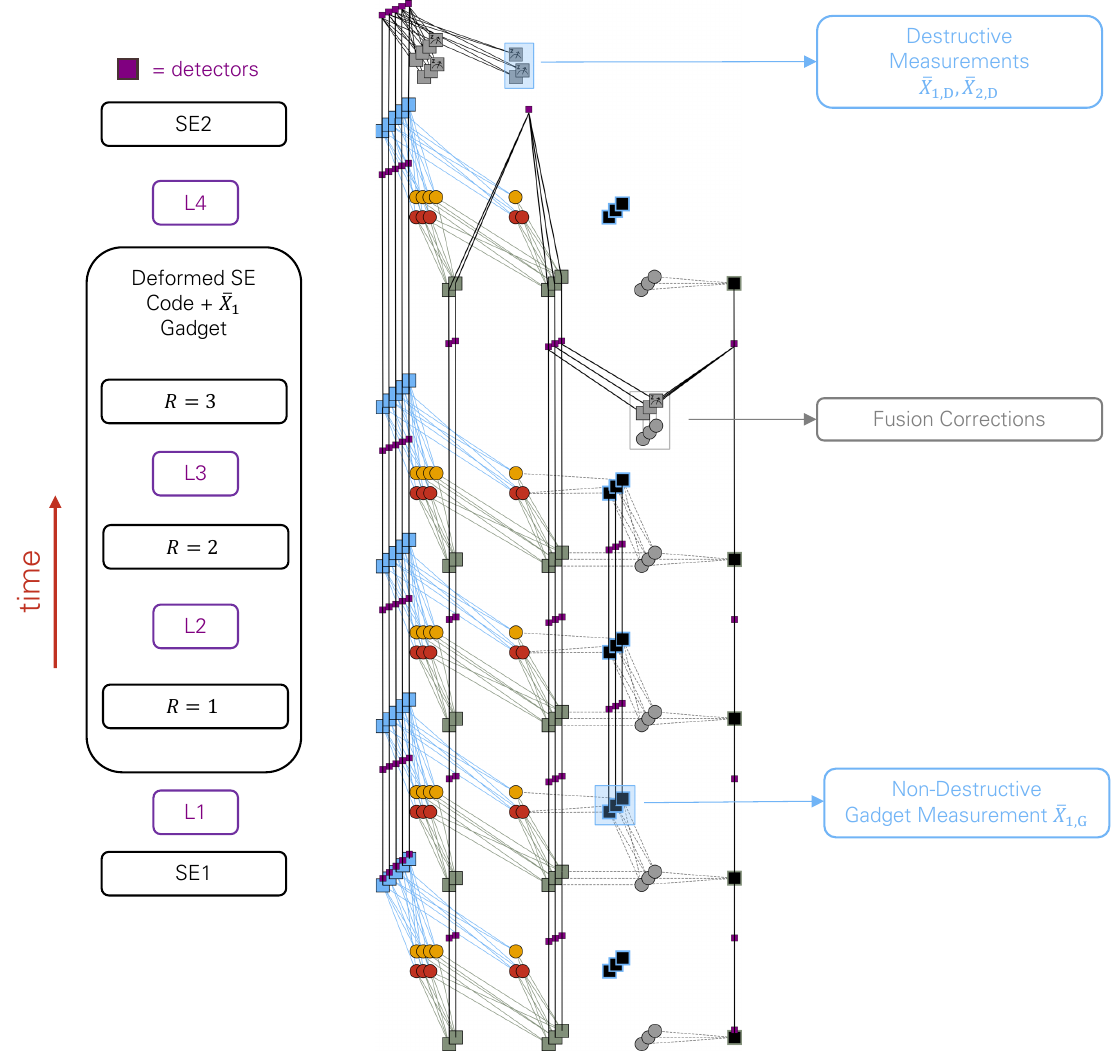}
    \caption{Detector matching graph of the 62 detectors for a three round X gadget experiment - detectors are defined on the bulk parities of deformed code and gadget checks, $\chi$ and $\gamma$ repetitions across rounds, and the $\kappa$ fusion parities when the gadget detaches.}
    \label{fig:gadget_3_round}
\end{figure*}

In \cref{fig:gadget_3_round}, we present the detector matching graph and space-time diagram of the three round $\bar{X}_{1,G}$ gadget experiment. 

\section{Simulation Methodology\label{app:b}}

In this section, we describe our methodology for converting a Stim circuit of each experiment into a circuit that can be executed on Quantinuum H2 hardware. 

We first create each circuit in Stim with a gate set of \{\texttt{Measure}, \texttt{Reset}, \texttt{H}, \texttt{X}, \texttt{CNOT}\}. We then use the PyTKET library to rebase and optimise the circuit to be compatible with Quantinuum H2 hardware. The circuit compiled to the H2-2 device consists of \{\texttt{Measure}, \texttt{Reset}, \texttt{Rx}, \texttt{Rz}, \texttt{ZZPhase(0.5)}\} operations with rotations of multiples of $\pi/2$ angles, hence this is still a Clifford circuit. The circuit is optimised using TKET optimisation level 2.

Finally, we use \texttt{Q\#} to convert to Quantum Intermediate Representation (QIR). This is sent to Azure with the following parameters specified: \texttt{no-opt} and \texttt{no-reduce} to disable further TKET or compiler optimisations and \texttt{leakage\_repump\_enable} to actively suppress leakage. We convert the compiled PyTKET circuit to a like-for-like Stim circuit and verify that the optimisation do not affect the circuit distance of the observables. 

\section{Error Model\label{app:c}}
In \cref{tab:error-model}, we present the error model used for simulations and detector error models in this work based on the performance validation data for the H2-2 device \cite{quantinuum_performance_validation}. We combine the figures for gate induced leakage with the respective gate error, as we enable the setting for leakage repump \cite{Hayes2020} converting leakage to Pauli-error. We do not model measurement cross-talk as the validation data quotes a value for cross-talk on neighbouring qubits and we lack low-level ion routing and operation batching data - we expect this not to be leading order as the cross-talk error is an order of magnitude below the next smallest error. Lacking exact operation batching details we estimate serial 2Q depth (the number of distinct instances of 2Q gates) by a greedy sorting of the gates into the 4 gate zones, and scaling the idle error per 1-layer circuit time accordingly. Data sheet at Ref. \cite{quantinuum_h2_datasheet}. 

\begin{figure*}
\centering
\includegraphics[width=\linewidth]{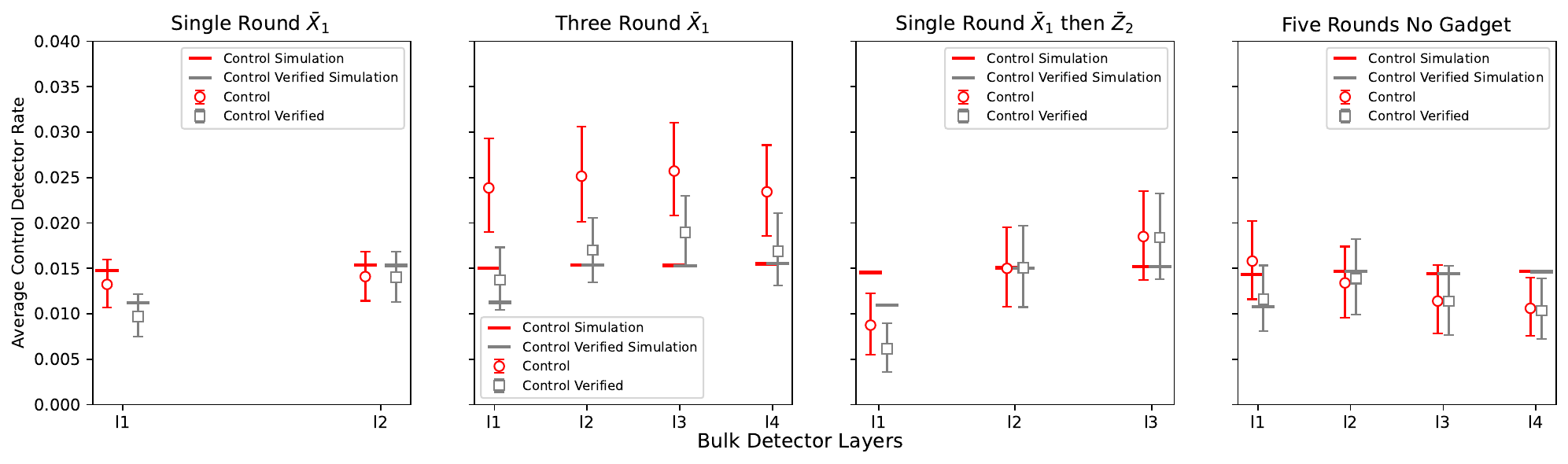}
\caption{Detector rate averaged over the control stabilisers against simulation under parameters of \cref{tab:error-model}. Error bars are bootstrapped 95\% intervals. Values counting all shots are red with circular markers, and verified shots are grey with square markers. Markers show experimental values and horizontal bars show simulated values. } \label{fig:control_group}
\end{figure*}

The control set detector rates are shown in Fig. \ref{fig:control_group}. Averaged over all layers and circuits the simulated detector rate is $1.049\times$ the measured rate. Comparing simulation to the average over layers in each circuit the simulated rates are $\left(1.10,0.92,1.06,1.14\right)\times$ the measured rate for the (XR1, XR3, SEQ, MEM) circuits respectively. We find the error model in close agreement with the hardware on average. Despite agreeing in the average, the measured detector rates vary over the layers by more than the sampling error. As this variation is of similar magnitude in circuits with and without gadgets, we attribute it to low-level compilation details not exposed to the user. This motivates \cref{fig:excess} where we compare the excess of the deformed detector's rate over the control group and compare to simulation, under the assumption per-layer effects are common to the whole register. 

\begin{table*}[]
    \centering
    \caption{Error model derived from performance validation data for the H2-2 device \cite{quantinuum_performance_validation}. The gate error rates are summed with the respective gate's leakage rate as hardware leakage repump is enabled.}
    \label{tab:error-model}
    \renewcommand{\arraystretch}{1.5} 
    \begin{tabular}{@{}|c|c|c|@{}}
    \hline
    \textbf{Error Type}                 & \textbf{Description}  & \textbf{Error Rate (Based on H2-2)}     \\
    \hline
    \multicolumn{1}{|c|}{Reset Error}   & \multicolumn{1}{c|}{X error post Z-basis reset}    & \multicolumn{1}{c|}{$(9.4\times10^{-4})/2$} \\
    \multicolumn{1}{|c|}{1Q Gate Error} & \multicolumn{1}{c|}{1Q Depolarising}       & \multicolumn{1}{c|}{$3.4\times10^{-5}$} \\
    \multicolumn{1}{|c|}{2Q Gate Error} & \multicolumn{1}{c|}{2Q Depolarising}     & \multicolumn{1}{c|}{$1.0\times10^{-3}$} \\
    \multicolumn{1}{|c|}{Measurement Error} & \multicolumn{1}{c|}{X error pre Z-basis reset}  & \multicolumn{1}{c|}{$(9.4\times10^{-4})/2$} \\
    \multicolumn{1}{|c|}{Idling Error}  & \multicolumn{1}{c|}{Z error on all qubits, aggregated per round}       & \multicolumn{1}{c|}{$(1.2\times10^{-4})/7$ per 2Q  depth} \\
    \hline
    \end{tabular}
\end{table*}

\section{Experimental Results\label{app:d}}
Tables \ref{tab:results-x1r1},\ref{tab:results-x1r3},\ref{tab:results-x1z2r1},\ref{tab:results-mem} show the error rates for the full set of experimental shots alongside the subset passing verification. Corrected labels Tesseract decoded shots without verification, Post-Verification labels the same decoding applied to only the verified shots. Verification rejects (2.9\%, 4.3\%, 2.2\%, 3.4\%) of shots for the (XR1, XR3, SEQ, MEM) circuits respectively. For XR1, SEQ, MEM verification has no effect on the number of errors and the reduced shot count therefore means slightly higher error rates than for the Corrected column. For the XR3 circuit, verification removes all 10 gadget errors, all from within the flagged loss-of-coherence region from \cref{fig:shotquality}.

\begin{table*}[!t]
\centering
\caption{Results for the single round $\bar{X}_1$ gadget experiment (XR1). Bounds represent 95\% Bayesian credible intervals using a Jeffreys prior ($\alpha = 0.5, \beta = 0.5$)}
\label{tab:results-x1r1}
\begin{adjustbox}{width=0.95\textwidth}
\begin{tabular}{@{}|c|c|c|ccccc|ccccc|@{}}
\toprule
 &  &  & \multicolumn{5}{c|}{\textbf{Corrected}} & \multicolumn{5}{c|}{\textbf{Post-Verification}} \\ \cmidrule(l){4-13} 
\multirow{-2}{*}{\textbf{Data Preparation}} & \multirow{-2}{*}{\textbf{Measurement}} & \multirow{-2}{*}{\textbf{Relative to}} & \multicolumn{1}{c|}{\textbf{Errors}} & \multicolumn{1}{c|}{\textbf{Shots}} & \multicolumn{1}{c|}{\textbf{Error Rate}} & \multicolumn{1}{c|}{\textbf{Lower Bound}} & \textbf{Upper Bound} & \multicolumn{1}{c|}{\textbf{Errors}} & \multicolumn{1}{c|}{\textbf{Shots}} & \multicolumn{1}{c|}{\textbf{Error Rate}} & \multicolumn{1}{c|}{\textbf{Lower Bound}} & \textbf{Upper Bound} \\ \midrule
$\lvert++\rangle$ & X1G & $\lvert+\rangle$ & \multicolumn{1}{c|}{6} & \multicolumn{1}{c|}{500} & \multicolumn{1}{c|}{1.20\%} & \multicolumn{1}{c|}{0.50\%} & 2.46\% & \multicolumn{1}{c|}{6} & \multicolumn{1}{c|}{487} & \multicolumn{1}{c|}{1.23\%} & \multicolumn{1}{c|}{0.52\%} & 2.52\% \\ \midrule
$\lvert--\rangle$ & X1G & $\lvert-\rangle$ & \multicolumn{1}{c|}{1} & \multicolumn{1}{c|}{500} & \multicolumn{1}{c|}{0.20\%} & \multicolumn{1}{c|}{0.02\%} & 0.93\% & \multicolumn{1}{c|}{1} & \multicolumn{1}{c|}{484} & \multicolumn{1}{c|}{0.21\%} & \multicolumn{1}{c|}{0.02\%} & 0.96\% \\ \midrule
\rowcolor[HTML]{E8E8E8} 
X Basis & X1G & Prepared State & \multicolumn{1}{c|}{7} & \multicolumn{1}{c|}{1000} & \multicolumn{1}{c|}{0.70\%} & \multicolumn{1}{c|}{0.31\%} & 1.37\% & \multicolumn{1}{c|}{7} & \multicolumn{1}{c|}{971} & \multicolumn{1}{c|}{0.72\%} & \multicolumn{1}{c|}{0.32\%} & 1.41\% \\ \midrule
\rowcolor[HTML]{FFFFFF} 
$\lvert++\rangle$ & X1D & $\lvert+\rangle$ & \multicolumn{1}{c|}{0} & \multicolumn{1}{c|}{500} & \multicolumn{1}{c|}{0.00\%} & \multicolumn{1}{c|}{0.00\%} & 0.50\% & \multicolumn{1}{c|}{0} & \multicolumn{1}{c|}{487} & \multicolumn{1}{c|}{0.00\%} & \multicolumn{1}{c|}{0.00\%} & 0.51\% \\ \midrule
$\lvert--\rangle$ & X1D & $\lvert-\rangle$ & \multicolumn{1}{c|}{0} & \multicolumn{1}{c|}{500} & \multicolumn{1}{c|}{0.00\%} & \multicolumn{1}{c|}{0.00\%} & 0.50\% & \multicolumn{1}{c|}{0} & \multicolumn{1}{c|}{484} & \multicolumn{1}{c|}{0.00\%} & \multicolumn{1}{c|}{0.00\%} & 0.52\% \\ \midrule
\rowcolor[HTML]{E8E8E8} 
X Basis & X1D & Prepared State & \multicolumn{1}{c|}{0} & \multicolumn{1}{c|}{1000} & \multicolumn{1}{c|}{0.00\%} & \multicolumn{1}{c|}{0.00\%} & 0.25\% & \multicolumn{1}{c|}{0} & \multicolumn{1}{c|}{971} & \multicolumn{1}{c|}{0.00\%} & \multicolumn{1}{c|}{0.00\%} & 0.26\% \\ \midrule
\rowcolor[HTML]{FFFFFF} 
$\lvert++\rangle$ & X2D & $\lvert+\rangle$ & \multicolumn{1}{c|}{0} & \multicolumn{1}{c|}{500} & \multicolumn{1}{c|}{0.00\%} & \multicolumn{1}{c|}{0.00\%} & 0.50\% & \multicolumn{1}{c|}{0} & \multicolumn{1}{c|}{487} & \multicolumn{1}{c|}{0.00\%} & \multicolumn{1}{c|}{0.00\%} & 0.51\% \\ \midrule
$\lvert--\rangle$ & X2D & $\lvert-\rangle$ & \multicolumn{1}{c|}{0} & \multicolumn{1}{c|}{500} & \multicolumn{1}{c|}{0.00\%} & \multicolumn{1}{c|}{0.00\%} & 0.50\% & \multicolumn{1}{c|}{0} & \multicolumn{1}{c|}{484} & \multicolumn{1}{c|}{0.00\%} & \multicolumn{1}{c|}{0.00\%} & 0.52\% \\ \midrule
\rowcolor[HTML]{E8E8E8} 
X Basis & X2D & Prepared State & \multicolumn{1}{c|}{0} & \multicolumn{1}{c|}{1000} & \multicolumn{1}{c|}{0.00\%} & \multicolumn{1}{c|}{0.00\%} & 0.25\% & \multicolumn{1}{c|}{0} & \multicolumn{1}{c|}{971} & \multicolumn{1}{c|}{0.00\%} & \multicolumn{1}{c|}{0.00\%} & 0.26\% \\ \midrule
\rowcolor[HTML]{FFFFFF} 
$\lvert++\rangle$ & X1G & X1D & \multicolumn{1}{c|}{6} & \multicolumn{1}{c|}{500} & \multicolumn{1}{c|}{1.20\%} & \multicolumn{1}{c|}{0.50\%} & 2.46\% & \multicolumn{1}{c|}{6} & \multicolumn{1}{c|}{487} & \multicolumn{1}{c|}{1.23\%} & \multicolumn{1}{c|}{0.52\%} & 2.52\% \\ \midrule
$\lvert--\rangle$ & X1G & X1D & \multicolumn{1}{c|}{1} & \multicolumn{1}{c|}{500} & \multicolumn{1}{c|}{0.20\%} & \multicolumn{1}{c|}{0.02\%} & 0.93\% & \multicolumn{1}{c|}{1} & \multicolumn{1}{c|}{484} & \multicolumn{1}{c|}{0.21\%} & \multicolumn{1}{c|}{0.02\%} & 0.96\% \\ \midrule
$\lvert00\rangle$ & X1G & X1D & \multicolumn{1}{c|}{5} & \multicolumn{1}{c|}{250} & \multicolumn{1}{c|}{2.00\%} & \multicolumn{1}{c|}{0.77\%} & 4.33\% & \multicolumn{1}{c|}{5} & \multicolumn{1}{c|}{240} & \multicolumn{1}{c|}{2.08\%} & \multicolumn{1}{c|}{0.80\%} & 4.51\% \\ \midrule
$\lvert11\rangle$ & X1G & X1D & \multicolumn{1}{c|}{3} & \multicolumn{1}{c|}{250} & \multicolumn{1}{c|}{1.20\%} & \multicolumn{1}{c|}{0.34\%} & 3.17\% & \multicolumn{1}{c|}{3} & \multicolumn{1}{c|}{245} & \multicolumn{1}{c|}{1.22\%} & \multicolumn{1}{c|}{0.35\%} & 3.23\% \\ \midrule
\rowcolor[HTML]{E8E8E8} 
X/Z Basis & X1G & X1D & \multicolumn{1}{c|}{15} & \multicolumn{1}{c|}{1500} & \multicolumn{1}{c|}{1.00\%} & \multicolumn{1}{c|}{0.59\%} & 1.60\% & \multicolumn{1}{c|}{15} & \multicolumn{1}{c|}{1456} & \multicolumn{1}{c|}{1.03\%} & \multicolumn{1}{c|}{0.60\%} & 1.65\% \\ 
\bottomrule
\end{tabular}
\end{adjustbox}
\end{table*}    

\begin{table*}[!t]
\centering
\caption{Results for the three round $\bar{X}_{1,G}$ circuit (XR3). Bounds represent 95\% Bayesian credible intervals using a Jeffreys prior ($\alpha = 0.5, \beta = 0.5$)}
\label{tab:results-x1r3}
\begin{adjustbox}{width=0.95\textwidth}
\begin{tabular}{@{}|c|c|c|ccccc|ccccc|@{}}
\toprule
 &  &  & \multicolumn{5}{c|}{\textbf{Corrected}} & \multicolumn{5}{c|}{\textbf{Post-Verification}} \\ \cmidrule(l){4-13} 
\multirow{-2}{*}{\textbf{Data Preparation}} & \multirow{-2}{*}{\textbf{Measurement}} & \multirow{-2}{*}{\textbf{Relative to}} & \multicolumn{1}{c|}{\textbf{Errors}} & \multicolumn{1}{c|}{\textbf{Shots}} & \multicolumn{1}{c|}{\textbf{Error Rate}} & \multicolumn{1}{c|}{\textbf{Lower Bound}} & \textbf{Upper Bound} & \multicolumn{1}{c|}{\textbf{Errors}} & \multicolumn{1}{c|}{\textbf{Shots}} & \multicolumn{1}{c|}{\textbf{Error Rate}} & \multicolumn{1}{c|}{\textbf{Lower Bound}} & \textbf{Upper Bound} \\ \midrule
\rowcolor[HTML]{FFFFFF} 
$\lvert++\rangle$ & X1G & $\lvert+\rangle$ & \multicolumn{1}{c|}{10} & \multicolumn{1}{c|}{1000} & \multicolumn{1}{c|}{1.00\%} & \multicolumn{1}{c|}{0.52\%} & 1.77\% & \multicolumn{1}{c|}{0} & \multicolumn{1}{c|}{957} & \multicolumn{1}{c|}{0.00\%} & \multicolumn{1}{c|}{0.00\%} & 0.26\% \\ \midrule
$\lvert++\rangle$ & X1D & $\lvert+\rangle$ & \multicolumn{1}{c|}{5} & \multicolumn{1}{c|}{1000} & \multicolumn{1}{c|}{0.50\%} & \multicolumn{1}{c|}{0.19\%} & 1.09\% & \multicolumn{1}{c|}{1} & \multicolumn{1}{c|}{957} & \multicolumn{1}{c|}{0.10\%} & \multicolumn{1}{c|}{0.01\%} & 0.49\% \\ \midrule
$\lvert++\rangle$ & X2D & $\lvert+\rangle$ & \multicolumn{1}{c|}{12} & \multicolumn{1}{c|}{1000} & \multicolumn{1}{c|}{1.20\%} & \multicolumn{1}{c|}{0.66\%} & 2.02\% & \multicolumn{1}{c|}{2} & \multicolumn{1}{c|}{957} & \multicolumn{1}{c|}{0.21\%} & \multicolumn{1}{c|}{0.04\%} & 0.67\% \\ \midrule
$\lvert++\rangle$ & X1G & X1D & \multicolumn{1}{c|}{11} & \multicolumn{1}{c|}{1000} & \multicolumn{1}{c|}{1.10\%} & \multicolumn{1}{c|}{0.59\%} & 1.90\% & \multicolumn{1}{c|}{1} & \multicolumn{1}{c|}{957} & \multicolumn{1}{c|}{0.10\%} & \multicolumn{1}{c|}{0.01\%} & 0.49\% \\ 
\bottomrule
\end{tabular}
\end{adjustbox}
\end{table*}    

\begin{table*}[!t]
\centering
\caption{Results for the sequential $\bar{X}_{1,G}$ then $ \bar{Z}_{2,G}$ circuit (SEQ). Bounds represent 95\% Bayesian credible intervals using a Jeffreys prior ($\alpha = 0.5, \beta = 0.5$)}
\label{tab:results-x1z2r1}
\begin{adjustbox}{width=0.95\textwidth}
\begin{tabular}{@{}|c|c|c|ccccc|ccccc|@{}}
\toprule
 &  &  & \multicolumn{5}{c|}{\textbf{Corrected}} & \multicolumn{5}{c|}{\textbf{Post-Verification}} \\ \cmidrule(l){4-13} 
\multirow{-2}{*}{\textbf{Data Preparation}} & \multirow{-2}{*}{\textbf{Measurement}} & \multirow{-2}{*}{\textbf{Relative to}} & \multicolumn{1}{c|}{\textbf{Errors}} & \multicolumn{1}{c|}{\textbf{Shots}} & \multicolumn{1}{c|}{\textbf{Error Rate}} & \multicolumn{1}{c|}{\textbf{Lower Bound}} & \textbf{Upper Bound} & \multicolumn{1}{c|}{\textbf{Errors}} & \multicolumn{1}{c|}{\textbf{Shots}} & \multicolumn{1}{c|}{\textbf{Error Rate}} & \multicolumn{1}{c|}{\textbf{Lower Bound}} & \textbf{Upper Bound} \\ \midrule
$\lvert00\rangle$ & Z2G & $\lvert0\rangle$ & \multicolumn{1}{c|}{6} & \multicolumn{1}{c|}{500} & \multicolumn{1}{c|}{1.20\%} & \multicolumn{1}{c|}{0.50\%} & 2.46\% & \multicolumn{1}{c|}{6} & \multicolumn{1}{c|}{492} & \multicolumn{1}{c|}{1.22\%} & \multicolumn{1}{c|}{0.51\%} & 2.50\% \\ \midrule
$\lvert11\rangle$ & Z2G & $\lvert1\rangle$ & \multicolumn{1}{c|}{5} & \multicolumn{1}{c|}{500} & \multicolumn{1}{c|}{1.00\%} & \multicolumn{1}{c|}{0.38\%} & 2.18\% & \multicolumn{1}{c|}{5} & \multicolumn{1}{c|}{486} & \multicolumn{1}{c|}{1.03\%} & \multicolumn{1}{c|}{0.39\%} & 2.24\% \\ \midrule
\rowcolor[HTML]{E8E8E8} 
Z Basis & Z2G & Prepared State & \multicolumn{1}{c|}{11} & \multicolumn{1}{c|}{1000} & \multicolumn{1}{c|}{1.10\%} & \multicolumn{1}{c|}{0.59\%} & 1.90\% & \multicolumn{1}{c|}{11} & \multicolumn{1}{c|}{978} & \multicolumn{1}{c|}{1.12\%} & \multicolumn{1}{c|}{0.60\%} & 1.94\% \\ \midrule
\rowcolor[HTML]{FFFFFF} 
$\lvert00\rangle$ & Z2D & $\lvert0\rangle$ & \multicolumn{1}{c|}{0} & \multicolumn{1}{c|}{500} & \multicolumn{1}{c|}{0.00\%} & \multicolumn{1}{c|}{0.00\%} & 0.50\% & \multicolumn{1}{c|}{0} & \multicolumn{1}{c|}{492} & \multicolumn{1}{c|}{0.00\%} & \multicolumn{1}{c|}{0.00\%} & 0.51\% \\ \midrule
$\lvert11\rangle$ & Z2D & $\lvert1\rangle$ & \multicolumn{1}{c|}{0} & \multicolumn{1}{c|}{500} & \multicolumn{1}{c|}{0.00\%} & \multicolumn{1}{c|}{0.00\%} & 0.50\% & \multicolumn{1}{c|}{0} & \multicolumn{1}{c|}{486} & \multicolumn{1}{c|}{0.00\%} & \multicolumn{1}{c|}{0.00\%} & 0.52\% \\ \midrule
\rowcolor[HTML]{E8E8E8} 
Z Basis & Z2D & Prepared State & \multicolumn{1}{c|}{0} & \multicolumn{1}{c|}{1000} & \multicolumn{1}{c|}{0.00\%} & \multicolumn{1}{c|}{0.00\%} & 0.25\% & \multicolumn{1}{c|}{0} & \multicolumn{1}{c|}{978} & \multicolumn{1}{c|}{0.00\%} & \multicolumn{1}{c|}{0.00\%} & 0.26\% \\ \midrule
\rowcolor[HTML]{FFFFFF} 
$\lvert00\rangle$ & Z2G & Z2D & \multicolumn{1}{c|}{6} & \multicolumn{1}{c|}{500} & \multicolumn{1}{c|}{1.20\%} & \multicolumn{1}{c|}{0.50\%} & 2.46\% & \multicolumn{1}{c|}{6} & \multicolumn{1}{c|}{492} & \multicolumn{1}{c|}{1.22\%} & \multicolumn{1}{c|}{0.51\%} & 2.50\% \\ \midrule
$\lvert11\rangle$ & Z2G & Z2D & \multicolumn{1}{c|}{5} & \multicolumn{1}{c|}{500} & \multicolumn{1}{c|}{1.00\%} & \multicolumn{1}{c|}{0.38\%} & 2.18\% & \multicolumn{1}{c|}{5} & \multicolumn{1}{c|}{486} & \multicolumn{1}{c|}{1.03\%} & \multicolumn{1}{c|}{0.39\%} & 2.24\% \\ \midrule
\rowcolor[HTML]{E8E8E8} 
Z Basis & Z2G & Z2D & \multicolumn{1}{c|}{11} & \multicolumn{1}{c|}{1000} & \multicolumn{1}{c|}{1.10\%} & \multicolumn{1}{c|}{0.59\%} & 1.90\% & \multicolumn{1}{c|}{11} & \multicolumn{1}{c|}{978} & \multicolumn{1}{c|}{1.12\%} & \multicolumn{1}{c|}{0.60\%} & 1.94\% \\
\bottomrule
\end{tabular}
\end{adjustbox}
\end{table*}    

\begin{table*}[!t]
\centering
\caption{Results for the five cycle memory circuit with no gadget (MEM). Bounds represent 95\% Bayesian credible intervals using a Jeffreys prior ($\alpha = 0.5, \beta = 0.5$)}
\label{tab:results-mem}
\begin{adjustbox}{width=0.95\textwidth}
\begin{tabular}{@{}|c|c|c|ccccc|ccccc|@{}}
\toprule
 &  &  & \multicolumn{5}{c|}{\textbf{Corrected}} & \multicolumn{5}{c|}{\textbf{Post-Verification}} \\ \cmidrule(l){4-13} 
\multirow{-2}{*}{\textbf{Data Preparation}} & \multirow{-2}{*}{\textbf{Measurement}} & \multirow{-2}{*}{\textbf{Relative to}} & \multicolumn{1}{c|}{\textbf{Errors}} & \multicolumn{1}{c|}{\textbf{Shots}} & \multicolumn{1}{c|}{\textbf{Error Rate}} & \multicolumn{1}{c|}{\textbf{Lower Bound}} & \textbf{Upper Bound} & \multicolumn{1}{c|}{\textbf{Errors}} & \multicolumn{1}{c|}{\textbf{Shots}} & \multicolumn{1}{c|}{\textbf{Error Rate}} & \multicolumn{1}{c|}{\textbf{Lower Bound}} & \textbf{Upper Bound} \\ \midrule
\rowcolor[HTML]{FFFFFF} 
$\lvert++\rangle$ & X1D & $\lvert+\rangle$ & \multicolumn{1}{c|}{0} & \multicolumn{1}{c|}{500} & \multicolumn{1}{c|}{0.00\%} & \multicolumn{1}{c|}{0.00\%} & 0.50\% & \multicolumn{1}{c|}{0} & \multicolumn{1}{c|}{483} & \multicolumn{1}{c|}{0.00\%} & \multicolumn{1}{c|}{0.00\%} & 0.52\% \\ \midrule
$\lvert++\rangle$ & X2D & $\lvert+\rangle$ & \multicolumn{1}{c|}{0} & \multicolumn{1}{c|}{500} & \multicolumn{1}{c|}{0.00\%} & \multicolumn{1}{c|}{0.00\%} & 0.50\% & \multicolumn{1}{c|}{0} & \multicolumn{1}{c|}{483} & \multicolumn{1}{c|}{0.00\%} & \multicolumn{1}{c|}{0.00\%} & 0.52\% \\ \midrule
\rowcolor[HTML]{E8E8E8} 
$\lvert++\rangle$ & X1D/X2D & $\lvert+\rangle$ & \multicolumn{1}{c|}{0} & \multicolumn{1}{c|}{1000} & \multicolumn{1}{c|}{0.00\%} & \multicolumn{1}{c|}{0.00\%} & 0.25\% & \multicolumn{1}{c|}{0} & \multicolumn{1}{c|}{966} & \multicolumn{1}{c|}{0.00\%} & \multicolumn{1}{c|}{0.00\%} & 0.26\% \\ 
\bottomrule
\end{tabular}
\end{adjustbox}

\vspace{2ex} 
\rule{0pt}{10cm}
\end{table*}

\clearpage
\bibliography{bibliography}

\end{document}